\documentclass{article}
\usepackage[utf8]{inputenc}
\usepackage[margin=1in]{geometry}
\usepackage{hyperref,amsmath,amssymb,amsthm}
\usepackage{graphicx,xcolor,float,tikz}
\usepackage{subfig}
\usepackage{xspace}
\usepackage{wasysym}
\usepackage{enumitem}
\setlist{leftmargin=*}
\usepackage[multiple]{footmisc}

\newcommand{\Flip}{{\sffamily Flip}\xspace}
\newcommand{\ReCom}{{\sffamily ReCom}\xspace}
\newcommand{\NC}{\hbox{\sffamily Node Choice}\xspace}

\usepackage[noend]{algpseudocode}
\usepackage[ruled,vlined]{algorithm2e}

\title{Recombination:\\ A family of Markov chains for 
redistricting}
\author{Daryl DeFord, Moon Duchin, and Justin Solomon\thanks{Following the convention in mathematics,
author order is alphabetical.}}
\date{\today}

\begin{document}

\maketitle
  \tableofcontents

\newpage

\begin{abstract}
\emph{Redistricting} is the problem of partitioning a set of geographical units 
into a fixed number of districts, subject to a list 
of often-vague rules and priorities.
In recent years, the use of randomized methods to sample from the vast space of districting plans has
been gaining traction in courts of law for identifying partisan gerrymanders, and it is now emerging as a possible 
analytical tool for legislatures and independent commissions. In this paper, we set up redistricting as a graph partition problem and
introduce a new family of Markov chains called Recombination (or \ReCom)
on the space of graph partitions.  The main point of comparison will be 
the commonly used \Flip walk, which randomly changes the assignment label of a single
node at a time.  We  present evidence that \ReCom mixes efficiently, especially in contrast
to the slow-mixing \Flip, and provide experiments that demonstrate its qualitative behavior.  
We demonstrate the advantages of \ReCom on real-world data and explain both the challenges of the Markov chain approach and the analytical tools that it enables. We close with a short case study involving the Virginia House of Delegates.
\end{abstract}

\section{Introduction}

In many countries, geographic regions are divided into districts that elect political representatives,
such as when states are divided into districts that elect individual members to the U.S.\ House of Representatives.  The task of drawing district boundaries, or \emph{redistricting}, is fraught with technical, practical, and political challenges, and the ultimate choice of a districting plan has 
major consequences in terms of which groups are able to elect candidates of choice.
Even the best-intentioned map-drawers have a formidable task in drawing plans whose structure promotes basic
fairness principles set out in law and in public opinion.  Further complicating matters, agenda-driven redistricting makes it common 
for line-drawers to \emph{gerrymander}, or to design plans specifically skewing elections toward a preferred outcome,
such as favoring or disfavoring a political party, demographic group, or collection of incumbents.

The fundamental technical challenge in the study of redistricting is to contend with the sheer number of possible ways to construct
districting plans.  State geographies admit enormous numbers of divisions into contiguous districts; 
even when winnowing down to districting plans that satisfy criteria set forth by legislatures or commissions, the number remains
far too large to enumerate all possible plans in a state.  The numbers are 
easily in the range of googols rather than billions, as we will explain below.

Recent methods for analyzing and comparing districting plans attempt to do so by placing a plan 
in the context of valid alternatives---that is, those that cut up the same jurisdiction by the same rules and with the structural
features of the geography and the pattern of voting held constant.
Modern computational techniques can generate large and diverse \emph{ensembles} of comparison plans, even if building 
the full space of alternatives is out of reach. 
These ensembles contain \emph{samples} from the full space of plans, aiming to help compare a plan's properties to the range of possible designs.  
For them to provide a proper counterfactual, however, we need some assurance of representative sampling, i.e., 
drawing from a probability distribution that successfully reflects the rules and priorities articulated by redistricters.

In one powerful application, ensembles have been used to 
conduct 
\emph{outlier analysis}, arguing that 
a proposed plan has properties that are extreme outliers relative to the comparison statistics of an ensemble of 
alternative plans.  Such methods have been used in a string of recent legal challenges to partisan gerrymanders
(Pennsylvania, North Carolina,
Michigan, Wisconsin, Ohio), which were all successful at the district court or state supreme court level.  
Outliers also received a significant amount of discussion at the U.S.\ Supreme Court, but the 5--4 majority declared that it 
was too hard for a federal court to decide ``how much is too much" of an outlier.  
The method is very much alive not only in state-level legal challenges but as a screening step for the initial adoption of plans,
and we expect numerous states to employ ensemble analysis in 2021 when new plans
are enacted around the country.  
These  
methods can help clarify the influence of each individual state's 
political geography as well as the tradeoffs in possible rules and criteria.

The inferences that can be drawn from ensembles rely heavily on the distribution from which the ensembles are sampled.  To that end, \emph{Markov chain Monte Carlo}, or MCMC, methods offer strong underlying theory
and heuristics, in the form of mixing theorems and convergence diagonistics.  Drawing from this literature, the new Markov chains described here
pass several tests of quality, even though (as is nearly always the case in applications) rigorous
mixing time bounds are still out of reach.  

Below, we define a new family of random walks  called \emph{recombination} (or \ReCom) Markov chains on the space of partitions of a graph into a fixed number of connected subgraphs.
Recombination chains are designed for applications in redistricting.  Compared to past sampling methods applied in this context---most prominently a simpler \Flip walk that randomly changes the labeling of individual geographic units along district borders---\ReCom has many favorable properties that make it well suited to the study of 
redistricting.  Critically for reliability of MCMC-based analysis, we present evidence that \ReCom mixes efficiently 
to a distribution that comports with traditional districting criteria, with little or no parameter-tuning by the user.

\begin{figure}[!h]
    \centering
 \subfloat[Initial Partition]{\includegraphics[height=1.5in]{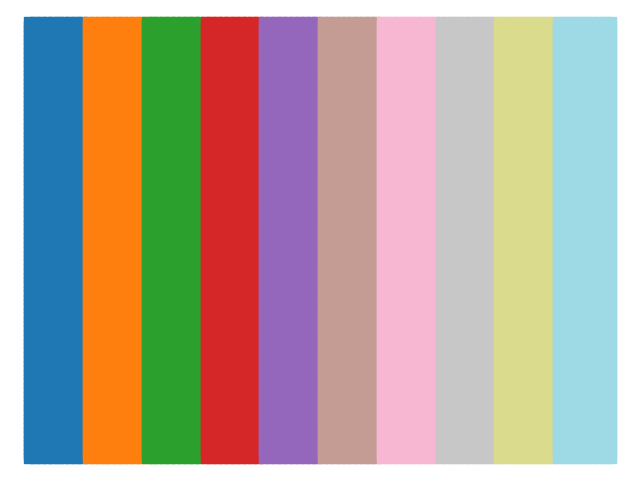}}\quad\ 
  \subfloat[1,000,000 \Flip steps]{\includegraphics[height=1.5in]{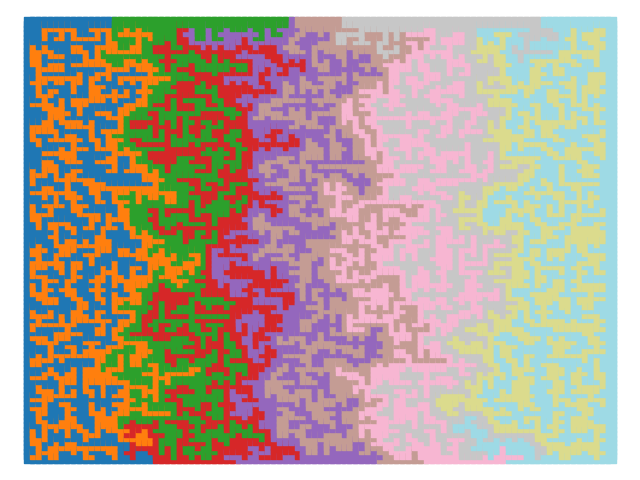}}\quad\
 \subfloat[100 \ReCom steps]{\includegraphics[height=1.5in]{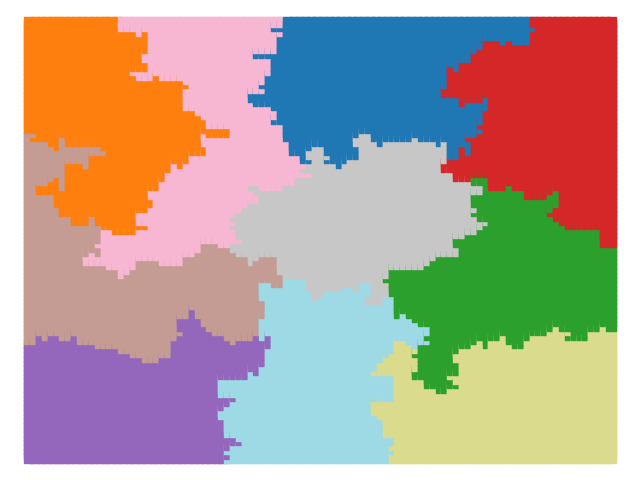}}    
    \caption{Comparison of the basic \Flip proposal versus the spanning tree \ReCom proposal to be described below. 
    Each Markov chain was run from the initial 
 partition of a $100\times100$ grid into 10 parts shown at left. }
    \label{fig:bchoice}
\end{figure}

\subsection{Novel contributions}
We introduce a new proposal distribution called \ReCom for MCMC on districting plans and argue that it 
provides an alternative to the previous \Flip-based approaches for ensemble-based analysis that makes
striking  improvements in efficiency and replicability.  In particular, we:
\begin{itemize}
    \item describe \ReCom and \Flip random walks on the space of graph partitions, discussing 
practical setup for implementing Markov chains for redistricting;
    \item provide evidence for fast mixing/stable results with \ReCom and contrast with properties of \Flip;
    \item study qualitative features of sampling distributions through experiments on real data,  addressing common 
    variants like simulated annealing and parallel tempering; \quad and
    \item provide a model analysis of racial gerrymandering in the Virginia House of Delegates.
\end{itemize}
To aid reproducibility of our work, an open-source implementation of \ReCom is available online \cite{gerrychain}. 
An earlier report on the Virginia House of Delegates written for reform advocates, legislators, and the general public 
is also available  \cite{VA-report}.

\subsection{Review of computational approaches to redistricting}\label{sec:computationalreview}

Computational methods for generating districting plans have appeared since at least the work of Weaver, Hess, and Nagel in the 1960s \cite{weaver1963procedure,nagel_simplified_1965}. 
Like much modern software for redistricting, early techniques like \cite{nagel_simplified_1965} incrementally improve districting plans in some metric while taking criteria like population balance, compactness, and partisan balance into account.  
Many basic  
elements still important for modern computational redistricting approaches were already in place in that work.
Quantitative criteria are extracted from redistricting practice (see our \S\ref{sec:operation});
contiguity is captured using a graph structure or ``touchlist'' (see our \S\ref{sec:redistrictingdiscrete});   a greedy hill-climbing strategy improves plans from an initial configuration; and randomization is used to improve the results.  
A version of the \Flip step (``the trading part'') even appears in their optimization procedure.
Their particular stochastic algorithm made use of hardware available at the time:  ``[R]un the same set of data cards a few times with the cards arranged in a different random order each time.''  

Since this initial exploration, computational redistricting has co-evolved with the development of modern algorithms and computing equipment.  Below, we highlight a few incomplete but representative examples; see \cite{cirincione_assessing_2000,tasnadi2011political,altman2010promise,saxon2018spatial,ricca2013political} for  broader surveys; only selected recent work is cited below.

\paragraph*{Optimization.}
Perhaps the most common redistricting approach discussed in the technical journal literature is the \emph{optimization} of districting plans.  Optimization algorithms are designed to extremize particular objective functions measuring plan properties while satisfying some set of constraints.  Most commonly, algorithms proposed for this task maintain contiguity and population balance of the districts and 
try to maximize the ``compactness" through some measure of shape \cite{kim2011optimization,jin2017spatial}.  
Many authors have used Voronoi or power diagrams with some variant
of $k$-means  \cite{fryer2011measuring,klein1,klein2,Levin2019AutomatedCR}, and there has been a lineage of approaches through integer programming \cite{buchanan} and even a partial 
differential equations approach with a volume-preserving curvature flow \cite{jacobs2018partial}.

Optimization algorithms have not so far 
become a significant element of reform efforts around redistricting practices, 
partly because of the difficulty of using them in assessment of proposed plans that take many other criteria into account. Moreover, most formulations of global optimization problems for full-scale districting plans are likely computationally intractable 
to solve, as most of the above authors acknowledge.

\paragraph*{Assembly.}
Here, a randomized process is used to create a plan from scratch, and this process is repeated to create a collection of plans
that will be used as a basis for comparison.  Note that an optimization algorithm with some stochasticity could be run 
repeatedly as an assembly algorithm, but generally the goals of assembly algorithms are to produce diversity where the goals
of optimization algorithms are to find a single best example.

The most basic assembly technique is to use a greedy {\em flood-fill} (agglomerative) strategy, starting from $k$ random choices
 among the geographical units as the seeds of districts and growing outwards by adding neighboring units until the 
 jurisdiction has been filled up and the plan is complete.  
Typically, these algorithms abandon a plan and re-start if they reach a configuration that cannot be completed into a valid plan, which can happen often. We are not aware of any theory to characterize the support and qualitative properties of the sampling distributions. 
Examples include \cite{chen_unintentional_2013, chen2016loser,magleby_new_2018}. 

\paragraph*{Random walks.}
A great deal of mathematical attention has recently focused on random walk approaches to redistricting.  
These methods use a step-by-step modification procedure to begin with one districting plan and incrementally 
transform it.  Examples include \cite{herschlag_quantifying_2018, herschlag_evaluating_2017, Fifield_A_2018,pegden,chikina2019practical}. 
An evolutionary-style variant with the same basic step
can be found in \cite{cho_toward_2016, liu_pear:_2016}.  
The use of random walks for sampling is well developed across scientific domains in the form 
of {\em Markov chain Monte Carlo}, or MCMC, techniques.  This is what the bulk of the present paper will consider.

We emphasize that while many of the techniques used in litigation are flip-based, they always involve customizations,
such as carefully tuned constraints and weighting, crossover steps, and so on.  The experiments below are not 
claimed to reproduce the precise setup of any of these  implementations (not least because the detailed specifications and
code are 
often not made public). Many of the drawbacks, limitations, and subtleties of working with flip chains  are well known to 
 practitioners but not yet present in the journal literature.  
In addition to discussing these aspects of flip chains,  we present an alternative chain that gives us an occasion to debate the mathematical 
and the math-modeling needs of the application to redistricting.

 \section{Markov chains}
\label{sec:Markov}

A Markov chain is simply a process for moving between positions in a {\em state space} according to a transition rule
in which the probability of arriving at a particular position at time $n+1$ depends only on the position at time $n$.  That is,
it is a random walk without memory.  A basic but powerful example of a Markov chain is the simple random walk 
on a graph:  from any node, the process chooses a neighboring node uniformly at random for the next step.  
More generally, one could take a weighted random walk on a graph, imposing different probabilities on the incident edges.
One of the fundamental facts in Markov chain theory is that any Markov chain can be accurately modeled as 
a (not necessarily simple) random walk on a (possibly directed) graph.  
Markov chains are used for a huge variety of applications, from Google's PageRank algorithm to speech recognition
to modeling phase transitions in physical materials.  
In particular, MCMC is a class of statistical methods that are used for sampling, with 
 a huge and fast-growing literature and a long track record of modeling success, including in a range of 
social science applications. See the classic survey \cite{Diaconis} for definitions, an introduction to Markov chain theory, and a lively guide to applications.

The theoretical appeal of Markov chains comes from the convergence guarantees that they provide. 
The fundamental theorem says that for any ergodic Markov chain there exists a unique stationary distribution, and that
iterating the transition step causes any initial state or probability distribution to converge to that steady state.  
The number of steps that it takes to pass a threshold of closeness to the steady state is called the {\em mixing time};
in applications, it is extremely rare to be able to rigorously prove  a bound on mixing time; 
instead, scientific authors often appeal to a suite of heuristic convergence tests.

This paper is devoted to investigating Markov chains for a {\em global} exploration of the universe of valid redistricting plans.
From a  mathematical perspective, the gold standard would be to define  Markov chains for which we can ($1$) characterize the stationary distribution $\pi$ and ($2$) compute the mixing time. 
(In most scientific applications, the stationary distribution is specified in advance through
the choice of an objective function and a Metropolis--Hastings or Gibbs sampler that weights states according to their scores.)
From a practical perspective in redistricting, confirming 
mixing to a distribution with a simple closed-form description is neither necessary nor sufficient.
For the application, a gold standard might be ($1'$) explanation of the distributional 
design and the weight that it places on particular kinds of districting plans, matched to the law and practice of redistricting,
and ($2'$) convergence heuristics and sensitivity analysis 
that give researchers confidence in the robustness and replicability of their techniques.

Stronger sampling and convergence theorems are available for {\em reversible} Markov
chains:  those for which the steady-state probability of being at state $P$ and transitioning to $Q$ equals the probability of 
being at $Q$ and transitioning to $P$ for all pairs $P,Q$ from the state space.  
In particular, a sequence of elegant theorems from the 1980s to now (Besag--Clifford \cite{besag1989generalized}, Chikina et al \cite{pegden,chikina2019practical}) shows that samples from reversible Markov chains admit conclusions about their likelihood of having been drawn from a
stationary distribution $\pi$ long before the sampling distribution approaches $\pi$.  
For redistricting, this theory enables what we might call {\em local} search:  while only sampling a relatively small 
neighborhood, we 
can draw conclusions about whether a plan has properties that are typical of random draws from $\pi$.
Importantly, these techniques can circumvent the mixing and convergence issues, but they must still contend
with issues of distributional design and sensitivity to user choice.

All previous MCMC methods we have encountered (for both local and global sampling) 
are built on variations of the same proposal distribution that we call a ``flip step," for which each move 
reassigns a single geographic unit from one district to a neighboring district.
This kind of proposal, for which we record several versions collectively denoted as \Flip, is relatively straightforward to implement and in its simplest form satisfies the properties needed for Markov chain theory to apply.  We will elaborate serious disadvantages of basic \Flip chains, however, in an attempt to catch the literature up with practitioner knowledge:   demonstrably slow mixing;
stationary distributions with undesirable qualitative properties; and additional complications in 
response to standard MCMC variations like constraining, re-weighting, annealing, and tempering.
We will argue that an alternative Markov chain design we call \emph{recombination}, implemented with a spanning tree step, avoids these problems.  We denote this alternative chain by \ReCom.
Both \Flip and \ReCom are discussed in detail in \S\ref{sec:proposals} below.

In applications, MCMC runs are often carried out with burn time (i.e., discarding the first $m$ steps) and subsampling 
(collecting every $r$ samples after that to create the ensemble).  If $r$ is set to match the mixing time, then 
the draws will be uncorrelated and the ensemble will be distributed according to the steady-state measure.
Experiments below will explore the choice of a suitable design for a \Flip chain.  
Some of the performance obstructions described below have led researchers to use extremely fast and/or parallelized
implementations, serious computing (or supercomputing) power, and various highly tuned or hybrid techniques that sometimes 
sacrifice the Markov property entirely or make external replicability impossible.  

On full-scale problems, a \ReCom chain with run length in the tens of thousands of steps produces ensembles
that pass many tests of quality, both in terms of convergence and in distributional design.  Depending on the details of the data, this can be run in anywhere from hours to a few days on a standard laptop.  

\section{Setting up the redistricting problem}

Before providing the technical details of \Flip and \ReCom, we set up the analysis of districting plans as a \emph{discrete} problem and explain how Markov chains can be designed to produce plans that comply with the rules of redistricting.

\subsection{Redistricting as a graph partition problem}\label{sec:redistrictingdiscrete}

The earliest understanding of pathologies that arise in redistricting was largely contour-driven.  Starting with the original ``gerrymander,'' whose salamander-shaped boundary inspired a famous 1812 political cartoon, irregular district boundaries on a map were understood to be signals that unfair division had taken place.
Several contemporary authors now argue 
for replacing the focus on contours with a discrete model \cite{duchin-tenner}, and in practice
the vast majority
of algorithmic approaches discussed above adopt the discrete model.
There are many reasons for this shift in perspective.  In practice, a district is an aggregation of a finite number of 
census blocks (defined by the Census Bureau every ten years) or precincts (defined by state, county, or local authorities).  
District boundaries extremely rarely 
cut through census blocks and typically preserve precincts,\footnote{For example,
the current Massachusetts plan splits just 1.5\% of precincts.  But measuring the degree of precinct preservation is very
difficult in most states because there is no precinct shapefile publicly available.} making it reasonable to compare a proposed plan to alternatives built from block or precinct units.  Furthermore, these discretizations give ample granularity; for instance, 
most states have five to ten thousand precincts and several hundred thousand census blocks.  

\begin{figure}[!h]
    \centering
\includegraphics[width=\textwidth]{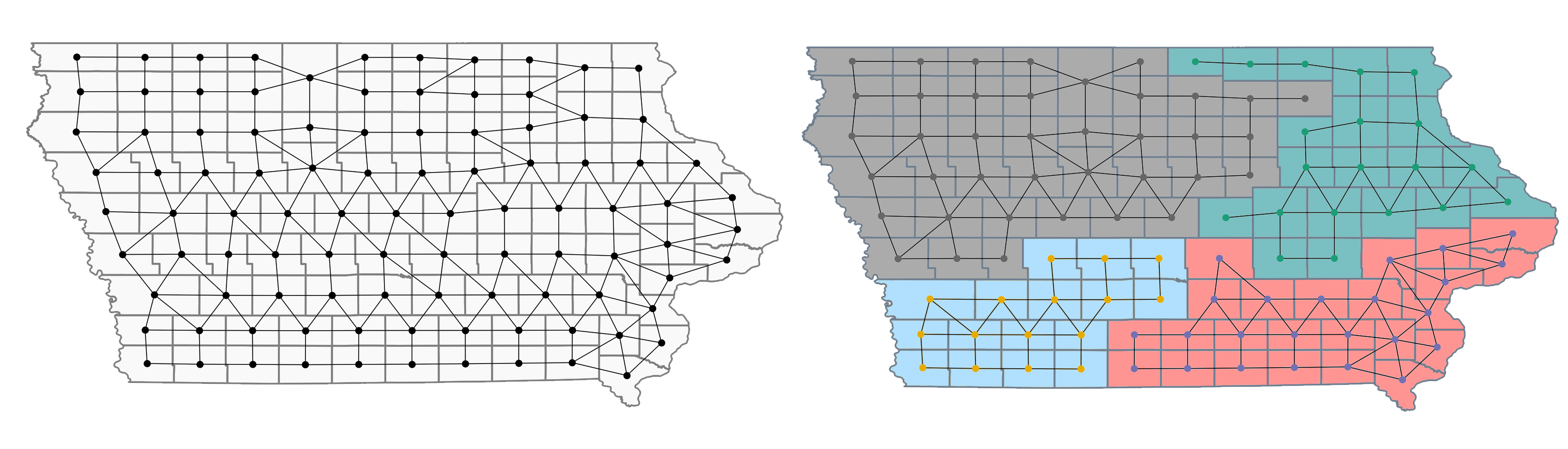}
    \caption{Iowa is currently the only state whose congressional districts are made of whole counties.
The dual graph of Iowa's counties is shown here together with the current Iowa congressional districts.  }
    \label{fig:dualgraph}
\end{figure}

From the discrete perspective, our basic object is the  {\em dual graph} to a geographic partition of the state into units.  
We build this graph $G=(V,E)$  by designating
 a vertex for each geographic unit (e.g., block or precinct) and placing edges in $E$ between those units that are geographically adjacent; Figure~\ref{fig:dualgraph} shows an example of this construction on the counties of Iowa.  
 With this formalism, a districting plan is a partition of the nodes of $V$ into subsets that induce connected components
 of $G$.  This way, redistricting can be understood as an instance of graph partitioning, a well-studied problem in combinatorics, applied math, and network science  \cite{NASCIMENTO,Schaeffer}. 
Equivalently, a districting plan is  an assignment of each node to a district via a labeling map $V\to \{1,\dots,k\}$.
The nodes (and sometimes the edges) of $G$ are decorated with assorted data, especially the population 
associated to each vertex, which is crucial for plan validity.  Other attributes for vertices may include the assignment of 
the unit to a municipality or a vector of its demographic statistics.  Relevant attributes attached to edges  might 
include the length of the boundary shared between the two adjacent units.

\subsubsection{Generating seed plans}
To actually run our Markov chains, we need a valid initial state---or \emph{seed}---in addition to the proposal method. 
Although in some situations we may want to start chains from the currently enacted plan, 
we will need other seed plans 
if we want to demonstrate that our ensembles are adequately independent of starting point.
Thus, it is useful to be able to construct starting plans that are at least contiguous and tolerably population balanced. 
Flood-fill methods (see \S\ref{sec:computationalreview} above) or spanning tree methods (see \S\ref{sec:recom} below) can be used 
for plan generation, and both are implemented in our codebase.

\subsection{Sampling  from the space of valid plans}
\label{sec:svp}

Increasing availability of computational resources has fundamentally changed the analysis and design of districting plans
by making it possible to explore the space of valid districting plans much more efficiently and fully.
It is now clear that any literal reading of the requirements governing redistricting permits an enormous number of potential plans for each state, far too many to build by hand or to consider systematically. The space of valid plans only grows if we account for the many possible readings of the criteria.

To illustrate this, consider the redistricting rule present in ten states that dictates that state House districts should nest
perfectly inside state Senate districts, either two-to-one (AK, IA, IL, MN, MT, NV, OR, WY) or three-to-one (OH, WI).  
A 
constraining way to interpret this mandate would be to fix the House districts
in advance and admit only those Senate plans that group appropriate numbers of adjacent House districts.  
Even under this 
narrow interpretation, a perfect matching analysis indicates that there are still 
6,156,723,718,225,577,984, or over $6\times 10^{18}$, ways to form valid state Senate plans just by pairing the current
House districts in Minnesota \cite{Alaska}.
The actual choice left to redistricters, who in reality control House and Senate lines simultaneously, is far more open,
and $10^{100}$ would not be an unreasonable guess.

\subsubsection{Operationalizing the rules}
\label{sec:operation}

 Securing \emph{operational} versions of rules and priorities governing the redistricting process requires a sequence of modeling decisions, with major consequences for the properties of the ensemble.
Constitutional and statutory provisions governing redistricting are never precise enough to admit a single unambiguous mathematical interpretation. 
We briefly survey the operationalization of important redistricting rules.
\begin{itemize}
    \item \textbf{Population balance:}   For each district,
we  can limit its  percentage deviation from the ideal size (state population divided by $k$, the number of districts).%
\footnote{The case law around tolerated population deviation is thorny and still evolving \cite[Chapter 1]{Realists}.  
For years, the basis of apportionment has been the raw
    population count from the decennial Census, but there are clear moves to change to a more restrictive population
    basis, such as by citizenship.}\footnote{
    Excessively tight requirements for population balance can spike the rejection rate of the Markov chain and impede
its efficiency.  Even for Congressional districts, which are often balanced to near-perfect equality in enacted plans, a precinct-based ensemble with 
$\le 1\%$ deviation can still provide a good comparator, because those plans typically can be quickly tuned by a mapmaker 
at the block level without breaking their other measurable features.}
    \item \textbf{Contiguity:} Most states require district contiguity by law, and it is the standard practice even when not formally required.  But even contiguity has subtleties in practice, because of water, corner adjacency, and the presence of 
    disconnected pieces.   Unfortunately, contiguity must be handled by building and cleaning dual graphs for each state 
    on a case-by-case basis.  
    \item \textbf{Compactness:} Most states have a ``compactness" rule preferring regular district shapes, but very few attempt
a definition, and several of the attempted definitions are unusable.\footnote{There are several standard scores in litigation,
 especially an isoperimetric score (``Polsby-Popper") and a comparison to the circumscribed circle (``Reock"), each 
 one applied to single districts.
It is easy to critique these scores, which are readily seen to be underdefined, unstable, and inconsistent \cite{duchin-tenner, bar2019gerrymandering,barnes,deford_total_2018}. 
In practice, compactness is almost everywhere ruled by the eyeball test.}
We will handle it in a mathematically natural manner for a discrete model:  we count the number of {\em cut edges}
in a plan, i.e.,  the number of edges in the dual graph whose endpoints belong to different districts (see \S\ref{sec:theory}).
This gives a notion of the discrete perimeter of a plan, and it corresponds well
to informal visual standards of regular district shapes (the ``eyeball test" that is used in practice much more heavily
than any score). 
    \item \textbf{Splitting rules:} Many states express a preference for districting plans that ``respect" or ``preserve" areas that are larger than the basic units of the plan, such as counties, municipalities, and (also underdefined) {\em communities of interest}.
There is no consensus on best practices for quantifying the relationship of a plan to a sparse set of geographical boundary curves.
Simply counting the number of units split (e.g., counties touching more than one district), or employing an entropy-like splitting score, are two alternatives that have been used in prior studies  \cite{Mattingly,VA-criteria}
    \item \textbf{Voting Rights Act (VRA):} The Voting Rights Act of 1965 is standing federal law that requires districts to be drawn to provide
qualifying minority groups with the opportunity to elect candidates of choice. \cite[Ch3-5]{Realists}.
Here, a modeler might reasonably choose to create a comparator ensemble made up of new plans that provide at least
as many districts with a substantial minority share of voting age population as the previous plan.%
\footnote{Since the VRA legal test involves assessing ``the totality of the circumstances," including local histories of discrimination and patterns of racially polarized voting, this is extraordinarily difficult to model in a Markov chain.  
However, the 
percentage of a minority group in the voting age population is frequently used as a proxy. 
For instance, in Virginia, there are two current Congressional districts with over 40\% Black Voting Age Population, 
and a plausible comparator ensemble should contain many plans that preserve that property.}  
    \item \textbf{Neutrality:} Often state rules will dictate that certain considerations should not be taken into account
    in the redistricting process, such as partisan data or incumbency status.  This is easily 
    handled in algorithm design
    by not recording or inputting associated data, like election results or incumbent addresses.
\end{itemize}
Finally, most of these criteria are subject to an additional decision about
\begin{itemize}
   \item \textbf{Aggregation and combination:} Many standard metrics used to analyze districting plans (as described above) are computed on a district-by-district basis, without specifying a scheme to aggregate scores across districts 
   to make plans mutually comparable.%
\footnote{If for instance we use an $L^\infty$ or supremum norm to summarize the compactness scores of the individual districts, then
all but the worst district can be altered with no penalty.  Choosing $L^1$ or $L^2$ aggregation takes all scores into account,
but to some extent allows better districts to cover for worse ones.  Pegden has argued for $L^{-1}$ to heavily
penalize the worst abuses for scores measured on a $[0,1]$ scale\cite{pegden,pegden1}.}
A modeler with multiple objective functions must also decide whether to try to combine them into a fused 
objective function, whether to threshold them at different levels, how to navigate a Pareto front of possible trade-offs, and so on.  
\end{itemize}
Our discussion in \S\ref{sec:experiments} provides details of how we approached some of the decisions above in our experiments.

\section{The flip and recombination chains}
\label{sec:proposals}

\subsection{Notation}\label{sec:notation}
Given a dual graph $G=(V,E)$, a $k$--partition of $G$ is a collection of disjoint subsets $P=\{V_1, V_2 \ldots, V_k\}$ such that $\bigsqcup V_i = V$. The full set of $k$--partitions of $G$ will be denoted $\mathcal{P}_k(G)$.  An element of $\mathcal P_k(G)$
may also be called a districting plan, or simply a plan, and the block of the partition with vertex set $V_i$ is sometimes 
called the $i$th district of the plan.

We may abuse notation by using the same symbol $P$ to denote the labeling function
$P:V\to \{1,\dots,k\}$.
That is, $P(u) = i$ means that $u\in V_i$ for the plan $P$. 
In a further notational shortcut, we will sometimes write $P(u)=V_i$
to emphasize that the labels index districts.
This labeling function allows us to represent the set of cut edges in the plan as 
$\partial P = \{(u,v)\in E : \ P(u)\neq P(v)\}$. We denote the set of boundary nodes by
$\partial_V P = \{u\in e: e\in \partial P \}$.
In the dual graphs derived from real-world data, our nodes are weighted with populations or other demographic data, which we represent with functions $V\to\mathbb{R}$.

This notation allows us to efficiently express constraints on the districts. For example, contiguity can be enforced by requiring that the induced subgraph on each $V_i$ is connected.  The  cut edge count described above as a measure of compactness
is written $|\partial P|$.  A condition that bounds population deviation can be written as
$$(1-\varepsilon) \frac{\sum_V w(v) }{k} 
\leq |V_i| \leq (1+\varepsilon)\frac{\sum_V w(v) }{k}.$$
For a given analysis or experiment, once  the constraints have been set and fixed, 
we will make use of a function $C:\mathcal{P}_k(G)\mapsto \{\texttt{True}, \texttt{False}\}$ to denote the validity check.
This avoids cumbersome notation to make explicit all of the individual constraints.

We next turn to setting out proposal methods for comparison.  
A proposal method is a procedure for transitioning between states of $\mathcal P_k(G)$ according to a proposal 
distribution.  
Formally, each $X_P$ is a  $[0,1]^{ \mathcal{P}_k(G)}$-valued random variable with coordinates summing to one, describing the transition probabilities.
Since $\mathcal P_k(G)$ is a gigantic but finite state space, the proposal distribution can be viewed as a stochastic matrix  with rows and columns indexed by the states $P$, such that the $(P,Q)$ entry $X_P(Q)$ is the probability of transitioning from $P$ to $Q$ 
in a single move.  
The resulting process is a Markov chain:  each successive state is drawn according to $X_P$, where $P$ is the current state.
Since these matrices are far too large to build, we may prefer to think of the proposal distribution as a stochastic algorithm for modifying the assignment of some subset of $V$. This latter perspective does not require computing transition probabilities explicitly, but rather leaves them implicit in the stochastic algorithm for modifying a partition.

In this section, we introduce the main \Flip and \ReCom proposals analyzed in the paper and describe some of their qualitative properties. We also devote some attention to the  spanning tree method that we employ in our empirical analysis. 

\subsection{\Flip proposals}\label{sec:flipproposal}

\subsubsection{Node choice, contiguity, rejection sampling}

\begin{figure}[!h]
    \centering
    \subfloat[Sequence of four flip steps]{\includegraphics[width = 5in]{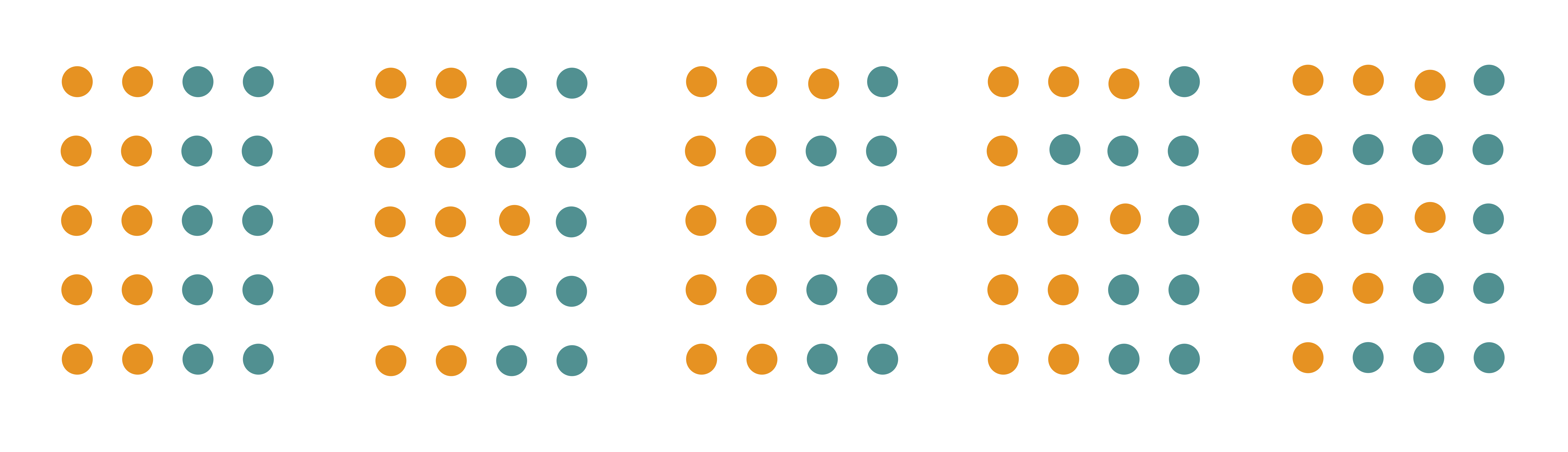}}\\
    \subfloat[Outcome of 500,000 flip steps]{\includegraphics[width=2.3in]{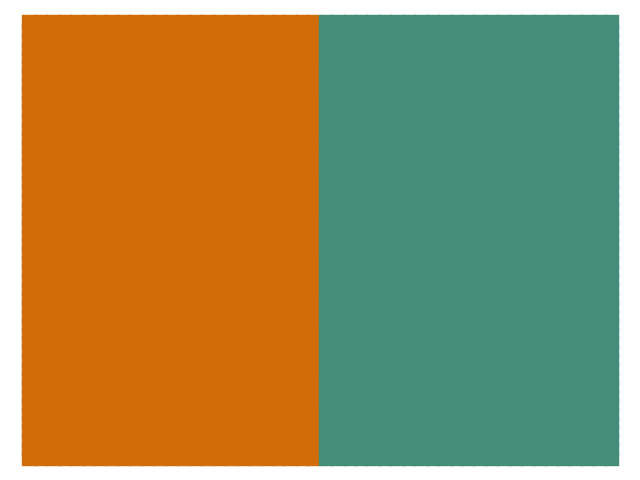}
\includegraphics[width=2.3in]{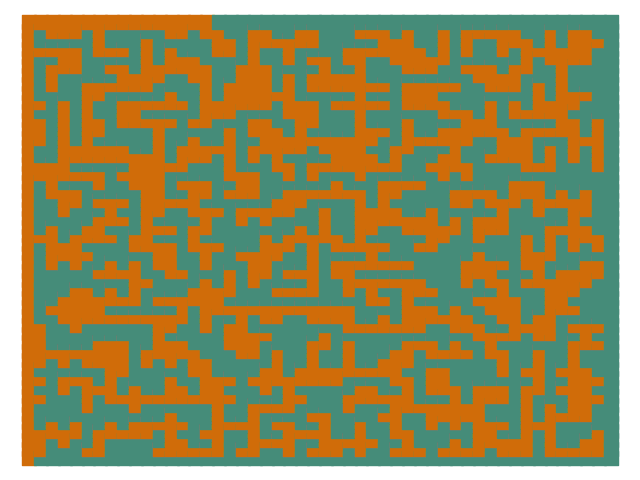}}
    \caption{At each flip step, a single node on the boundary changes assignment, preserving contiguity. 
This is illustrated schematically on a $5\times 4$ grid and then the end state of a long run is depicted on a $50\times 50$ grid.}
    \label{fig:flipschematic}
\end{figure}

At its simplest, a \Flip proposal changes the assignment of a single node at each step in the chain in a manner that
preserves the contiguity of the plan. See Figure \ref{fig:flipschematic} for a sequence of steps in this type of Markov chain and a randomly generated 2-partition of a $50\times 50$ grid, representative of the types of partitions generated by \Flip and its variants. This procedure provides a convenient vehicle for exploring the complexity of the partition sampling problem.

To implement  \Flip, we must decide how to select a node whose assignment will change, for which we define an intermediate process called \NC.
To ensure contiguity, it is intuitive to begin by choosing a vertex of $\partial_V P$ or an edge of $\partial P$, but because
degrees vary, this can introduce non-uniformity to the process.
To construct a {\em reversible} Markov chain we follow \cite{pegden} and instead sample uniformly from the set of (node, district) pairs $(u,V_i)$ where $u\in \partial_V P$ and there exists a cut edge $(u,v)\in \partial P$ with $P(v) = V_i$. This procedure amounts to making a  uniform choice among the partitions that differ only by the assignment of a single boundary node. Pseudocode for this method is presented in Algorithm \ref{alg:bflip}. 
The associated Markov chain has transition probabilities given by
$$X_P(Q) = \begin{cases}\frac1{|\{(v,P(w)):(v,w)\in \partial P \}|}& |\{P(u) \neq Q(u): u\in \partial P \}|=1\ \textrm{and}\ |\{P(u) \neq Q(u): u\notin \partial P \}|=0;\\
0&\textrm{otherwise}. 
\end{cases} $$ 
This can be interpreted as a simple random walk on $\mathcal{P}_k(G)$ where two partitions are connected if they differ at a single boundary node. Thus, the Markov chain is reversible.  Its stationary distribution is non-uniform, 
since each plan is weighted proportionally to the number of (node, district) pairs in its boundary. 

\noindent\begin{minipage}{\textwidth}
   \centering
   \begin{minipage}{.45\textwidth}
     \centering
     
\begin{algorithm}[H]
\caption{\texttt{Node Choice}}\label{alg:bflip}

\textbf{Input:} Dual graph $G=(V,E)$ and current partition $P$\\
\textbf{Output:} A new partition $Q$\\
\hfill \\
\SetAlgoLined
{\bf Select:} A (node, district) pair  $(u, V_i)$ uniformly from $\{(v,P(w)):(v,w)\in \partial P \}$\\
{\bf Define:} $Q(v)=\begin{cases} V_i  & \textrm{if}\ 
u = v\\
P(v)&\textrm{otherwise.}
\end{cases}$\\
\hfill \\
{\bf Return:} $Q$
\end{algorithm}
\end{minipage}
   \begin{minipage}{.45\textwidth}
     \centering
\begin{algorithm}[H]
\caption{\texttt{Flip}}\label{alg:flip}

\textbf{Input:} Dual graph $G=(V,E)$ and the current partition $P$\\
\textbf{Output:} A new partition $Q$\\
\hfill \\
\SetAlgoLined
{\bf Initialize:} \textit{Allowed} = \texttt{False}\\
\While{\textit{Allowed} = \texttt{False}}{
 $Q = \NC(G,P)$ \\
$\textit{Allowed} = C(Q)$\\
}
{\bf Return:} $Q$
\end{algorithm}
   \end{minipage}   
   \end{minipage}

At each step, the \NC algorithm grows one district by a node and shrinks another.  
One can quickly verify that a \NC step  maintains contiguity
in the district that grows but may break contiguity in the district that shrinks.  
In fact, after many steps it is likely to produce a plan with no contiguous districts at all.
To address this, we adopt a rejection sampling approach, only accepting contiguous proposals.
This produces our basic \Flip chain (see Algorithm \ref{alg:flip} for pseudocode and Figures \ref{fig:bchoice},\ref{fig:flipschematic} for visuals).
The rejection setup does not break reversibility of the associated Markov chain, since it now amounts
to a simple random walk on the restricted state space.

Rejection sampling is practical because it is far more efficient to evaluate whether or not a particular plan is permissible than to determine the full set of adjacent plans at each step. Both the size of the state space and the relatively expensive computations that are required at the scale of real-world dual graphs contribute to this issue. If the proposal fails contiguity or another constraint check, we simply generate  new proposed plans from the previous state until one
passes the check.

These methods have the advantage of explainability in court and step-by-step efficiency for computational purposes, since each new proposed plan is only a small perturbation of the previous one. 
The same property that allows this apparent computational advantage, however, 
also makes 
it difficult for \Flip-type proposals to explore the space of permissible plans efficiently. Figure \ref{fig:bchoice} shows that after 1 million steps the structure of the initial state is still clearly visible, and we will discuss evidence below that 1 billion steps is enough to improve matters significantly, but not to the point of mixing. Thus, the actual computational advantage is less clear, as it may take a substantially larger number of steps of the chain to provide reliable samples. 
This issue is exacerbated when legal criteria impose strict constraints on the space of plans, which may easily cause disconnectedness under this proposal.\footnote{A user can choose 
to ensure connectivity by relaxing even hard legal constraints during the run and winnowing to a valid sample later, which requires additional choices and tuning.}
Attempts have been made to address this mixing issue in practice with simulated annealing or parallel tempering \cite{herschlag_evaluating_2017,herschlag_quantifying_2018,Fifield_A_2018}, but  we will show in \S\ref{sec:experiments} 
that on the scale of real-world problems, these fixes do not immediately overcome the fundamental barrier to successful sampling
that is caused by the combination of extremely slow mixing and the domination of distended shapes.

\subsubsection{Uniformizing}

For practical interpretation of a sample, it can be useful to have a simple description of the sampling distribution.  
Although Algorithm \ref{alg:flip} does not have a uniform steady state distribution,  it is possible to re-weight the transition probabilities to obtain a {\em uniform} distribution, as in the work of Chikina--Frieze--Pegden \cite{pegden}. This can be done by adding self-loops to each plan in the state space to equalize the degree; the resulting technique is given in Algorithm \ref{alg:uflip}.
To see that this gives the uniform distribution over the permissible partitions of  $\mathcal{P}_k(G)$, we note that with  $M$ set to the maximum degree in the state space and
$p= \dfrac{|\{(u,P(v)) : (u,v) \in \partial P\}|}{M\cdot|V|}$ we have
$$X_P(Q) = \begin{cases} 1-p & Q=P\\
p & |\{P(u) \neq Q(u): u\in V \}|=1\\ 
0&\textrm{otherwise}.
\end{cases} $$ 
Continuing to follow 
Chikina et al, 
we can accelerate the \texttt{Uniform Flip} algorithm without changing its proposal distribution by employing a function that returns an appropriate number of steps to wait at the current state before transitioning, so as to simulate the  
expected self-loops traversed before a non-loop edge is chosen.
This variant is in Algorithm \ref{alg:uwflip}. Since the  geometric variable computes the expected waiting time before selecting
a node from $\partial_V P$,  this recovers the same walk and distribution with many fewer calls to the proposal function.

\noindent\begin{minipage}{\textwidth}
   \centering
   \begin{minipage}{.45\textwidth}
     \centering

 \begin{algorithm}[H]
\caption{\texttt{Uniform Flip}}\label{alg:uflip}

\textbf{Input:} Dual graph $G=(V,E)$ and current partition $P$\\
\textbf{Output:} New partition $Q$\\
\hfill \\
\SetAlgoLined
{\bf Initialize:} $p= \dfrac{|\{(u,P(v)): (u,v) \in \partial P\}|}{M\cdot|V|}$\\
\hfill\\

\eIf{\textrm{Bernoulli}(1-p) = 0}{
{\bf Return:} P}
{
\textit{Allowed} = \texttt{False}\\
\While{\textit{Allowed} = \texttt{False}}{
 $Q = \NC(G,p)$ \\
$\textit{Allowed} = C(Q)$\\
}
{\bf Return:} $Q$
}

\end{algorithm}
   \end{minipage}
   \begin{minipage}{.45\textwidth}
     \centering
     
\begin{algorithm}[H]
\caption{\texttt{Uniform Flip (Fast)}}\label{alg:uwflip}

\textbf{Input:} Dual graph $G=(V,E)$ and current partition $P$\\
\textbf{Output:} Number of steps to wait in the current state ($\sigma$) and  next partition ($Q$)\\
\hfill \\
\hfill\\
\SetAlgoLined
{\bf Initialize:} $p= \dfrac{|\{(u,P(v)): (u,v) \in \partial P\}|}{M\cdot|V|}$\\
\vspace{.42em}
$\sigma = \textrm{Geometric}(1-p)$
\hfill\\
\hfill\\
 $Q = \NC(G,p)$ \\
 \eIf{$C(Q) = \texttt{False}$}{{\bf Return:} $(\sigma, P)$\\}{{\bf Return:} $(\sigma, Q)$}
\end{algorithm}

   \end{minipage}

   \end{minipage}

\vspace{1em}

On the other hand, attempting to sample from the uniform distribution causes problems for at least two reasons. 
First, sampling from uniform distributions over partitions runs into complexity obstructions (see \S\ref{sec:RP=NP} below), implying as a corollary that we should not expect these chains to mix rapidly.  
And even though abstract complexity results do not always dictate practical performance, in practice we will observe that
extremely long runs are needed to produce substantial change in the map.
We will demonstrate slow mixing on problems approximating real scale by showing that even the projection to summary statistics remains strongly correlated with the initial state (Figure~\ref{fig:grid_plots}). 
Secondly, even if we had a uniform sampling oracle, the distribution is massively concentrated in non-compact plans:
generic connected partitions are remarkably snaky, with long tendrils and complex boundaries (see \S\ref{sec:fractal}).
The erraticness of typical shapes in the flip ensembles is undesirable from the perspective of districting, which places a premium on well-behaved boundaries. This also means that it is difficult for these chains to move effectively in the state space when compactness constraints are enforced, since generic steps increase the boundary length, leading to high rejection probabilities or disconnected state spaces. 
We evaluate some standard techniques for ameliorating this issue in our experiments below.
Correcting the shape problem is not straightforward and introduces a collection of parameters that interact in 
complicated ways with the other districting rules and criteria.

\subsection{\ReCom proposals}\label{sec:recom}

The slow mixing and poor qualitative behavior 
of the \Flip chain leads us to introduce a new Markov chain on partitions, which changes the assignment of many vertices at once while preserving contiguity.  Our new proposal is more computationally costly than \Flip at each step in the Markov chain, but this tradeoff is net favorable thanks to superior mixing and distributional properties.

\begin{figure}[!h]
    \centering
    \includegraphics[width = 5in]{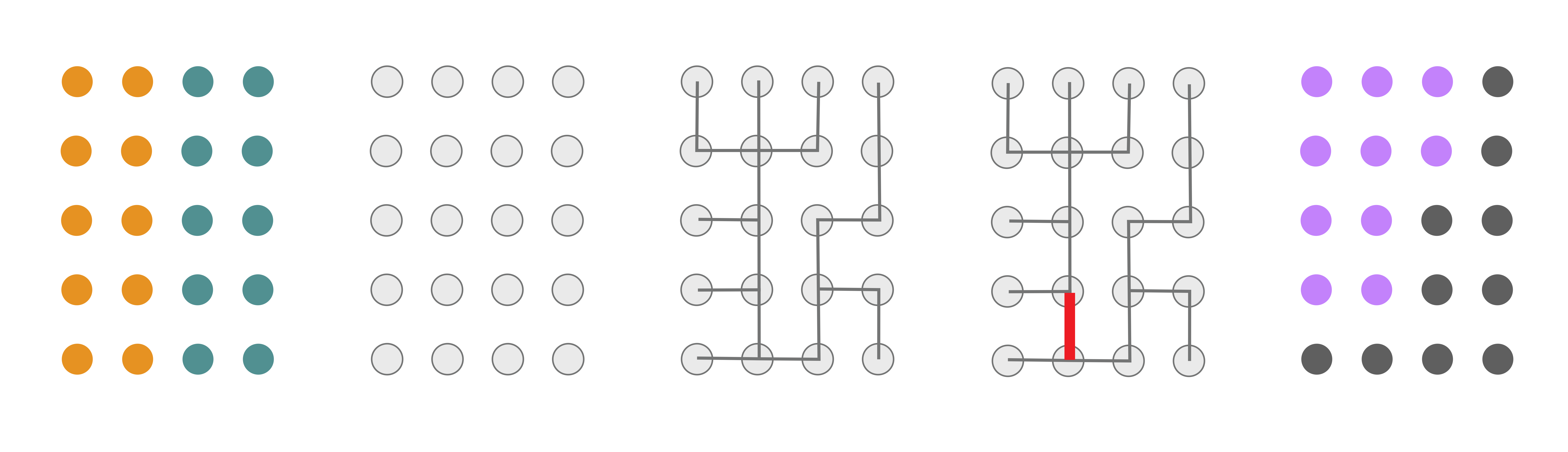}
    \caption{A schematic of a \ReCom spanning tree step for a small grid with $k=2$ districts. Deleting the indicated edge from the spanning tree leaves two connected components with an equal number of nodes.  }
    \label{fig:treerecom}
\end{figure}

At each step of our new chain, we select a number of districts of the current plan and form the induced subgraph of the dual graph on the nodes of those districts. 
We then partition this new region according to an algorithm that preserves contiguity of the districts.  We call this procedure {\em recombination} (\ReCom), motivated by the biological metaphor of recombining genetic information.  A general version of this approach is summarized in Algorithm \ref{alg:generalrecom}; Figure \ref{fig:treerecom}  shows a schematic of a single step with this proposal. 

\begin{algorithm}[!h]
\caption{\texttt{Recombination (General)}}\label{alg:generalrecom}
\textbf{Input:} Dual graph $G=(V,E)$, the current partition $P$, the number of districts to merge $\ell$\\
\textbf{Output:} The next partition $Q$\\
\hfill \\
\SetAlgoLined
 Select $\ell$  districts $W_1, W_2,\ldots, W_\ell$ from $P$.\\
Form the induced subgraph $H$ of $G$ on the nodes of $W=\bigcup_{i=1}^\ell W_i$.\\
Create a partition $R=\{U_1, U_2, \ldots, U_\ell\}$ of $H$\\
Define $Q(v) = \begin{cases} R(v) & \textrm{if}\ v\in H\\
P(v)&\textrm{otherwise}
\end{cases}$
\hfill\\
{\bf Return:} $Q$
\end{algorithm}

The \ReCom procedure in Algorithm~\ref{alg:generalrecom} is extremely general.  There are two algorithmic design decisions that are required to specify the details of a \ReCom chain:
\begin{itemize}
\item The first parameter in the \ReCom method is how to \textbf{choose which districts are merged} at each step. By fixing the partitioning method, we can create entirely new plans as in \S\ref{sec:redistrictingdiscrete} by merging all of the districts at each step ($\ell=k)$.
For most of our use cases, we work at the other extreme, taking two districts at a time ($\ell=2$), and we 
select our pair of adjacent districts to be merged proportionally to the  length of the boundary between them, which improves compactness quickly, as we will discuss in \S\ref{sec:fractal}.\footnote{Bipartitioning is usually easier to study than $\ell$-partitioning for $\ell>2$.  More importantly for this work,
the slow step in a recombination chain is the selection of a spanning tree.  Drawing spanning trees for the 
$\ell=k$ case (the full graph) is many times slower than for $\ell=2$ when $k$ is large, making bipartitioning a better 
choice for chain efficiency.
This approach also generalizes in a second way:  We can take a (maximal) matching on the dual graph of districts and bipartition each merged pair independently, taking advantage of the well-developed and effective theory of matchings.}  

\item The choice of \textbf{(re)partitioning method} offers more freedom. Desirable features include full support
over contiguous partitions, ergodicity of the underlying chain, ability to control the distribution with respect to legal features (particularly population balance), computational efficiency,  and ease of explanation in non-academic contexts like court cases and reform efforts. Potential examples include standard graph algorithms, like the spanning tree partitioning method we will introduce in \S\ref{sec:spanningtreerecom}, as well as methods based on minimum cuts, spectral clustering, or shortest paths between boundary points. 
\end{itemize}

With these two choices, we have a well-defined Markov chain to study.
The experiments shown in the present paper are conducted with a spanning tree method of bipartitioning,
which we now describe.

\subsubsection{Spanning tree recombination}\label{sec:spanningtreerecom}

In all experiments below, we focus on a particular method of bipartitioning that creates a recombination chain 
whose behavior is well-aligned to redistricting.  This method merges two adjacent districts, selects a spanning 
tree of the merged subgraph, and cuts it to form two new districts.

\begin{itemize}
    \item First, draw a spanning tree uniformly at random from among all of the spanning trees of the merged region.  Our implementation uses the loop-erased random walk method of Wilson's algorithm \cite{Wilson}.\footnote{Wilson's algorithm
is notable in that it samples uniformly from all possible spanning trees in polynomial time.}
    \item Next, seek an edge to cut that balances the population within the permitted tolerance. For an arbitrary spanning tree, it is not always possible to find such an edge, in which case we draw a new tree; this is another example of rejection sampling in our implementation. In practice, the rejection rate is low enough that this step runs efficiently. If there are multiple edges that could be cut to generate partitions with the desired tolerance, we sample uniformly from among them. 
\end{itemize}
Pseudocode for this technique is provided in Algorithm \ref{alg:treerecom}.

\begin{algorithm}[!h]
\caption{\texttt{ReCom} (Spanning tree bipartitioning)}\label{alg:treerecom}
\textbf{Input:} Dual graph $G=(V,E)$, the current partition $P$, population tolerance $\varepsilon$\\
\textbf{Output:} The next partition $Q$\\
\hfill \\
\SetAlgoLined
 {\bf Select:} $(u,v)\in \partial P$ uniformly\\
 Set $W_1=P(u)$ and $W_2=P(v)$\\
 Form the induced subgraph $H$ of $G$ on the nodes of $W_1\cup W_2$.\\
{\bf Initialize:} Cuttable = \texttt{False}\\
\While{Cuttable = False}{
Sample a spanning tree $T$ of $H$\\
Let EdgeList = []\\
\For{edge in T}{
Let $T_1, T_2 = T \setminus edge$ \\
\If{$|T_1|-|T_2|<\varepsilon|T|$}{Add edge to EdgeList\\ Cuttable = True\\ }
}
}
Select cut uniformly from EdgeList\\
Let $R= T \setminus cut$\\
Define $Q(v) = \begin{cases} R(v) & v\in H\\
P(v)&\textrm{otherwise}
\end{cases}$
\hfill\\
{\bf Return:} $Q$
\end{algorithm}

A similar spanning tree approach to creating initial seeds is available:  draw a spanning tree for the entire 
graph $G$, then recursively seek edges to cut that leave one complementary component of appropriate population
for a district.   

\section{Theoretical comparison}
\label{sec:theory}

Below, we will conduct experiments 
that provide intuition for qualitative behavior of the \Flip and \ReCom 
chains.  However, 
precise mathematical characterization of their stationary distributions appears to be extremely challenging and is the subject
of active research.  In this section, we provide high-level explanations of the two main phenomena that can 
be gleaned from experiments:  \ReCom samples preferentially
from fairly compact districting plans while simple \Flip ensembles are composed of plans with long and winding boundaries;
and \ReCom seems to mix efficiently while \Flip mixes slowly. 

\subsection{Distributional design: compactness}\label{sec:compactnessdesign}\label{sec:fractal}

``Compactness" is a vague but important  term in redistricting:   compact districts are those with 
tamer or plumper shapes.  
This can refer to having high area relative to perimeter, shorter boundary length, fewer spikes or necks or tentacles, 
and so on.  In this treatment, we focus on the discrete perimeter 
as a way to measure compactness.  
Recall from \S\ref{sec:notation} that 
for a plan $P$ that partitions a graph $G=(V,E)$, we  denote by $\partial P\subset E$ its set of cut edges, or the edges of $G$ whose endpoints are in different districts of $P$.  A slight variant is 
to count the number of boundary nodes $\partial_V P\subset V$ (those nodes at the endpoint of some cut edge).  
There is a great deal of mathematical literature connected to combinatorial perimeter, from the minimum cut problem 
to the Cheeger constant to expander graphs.  Though we focus on the discrete compactness scores here,
a dizzying array of compactness metrics has been proposed in connection to redistricting, and the analysis below---that \Flip is  must contend with serious compactness problems---would apply to any reasonable score, as the figures illustrate.  

\begin{figure}[!h]
    \centering
\subfloat{\includegraphics[height=1.25in]{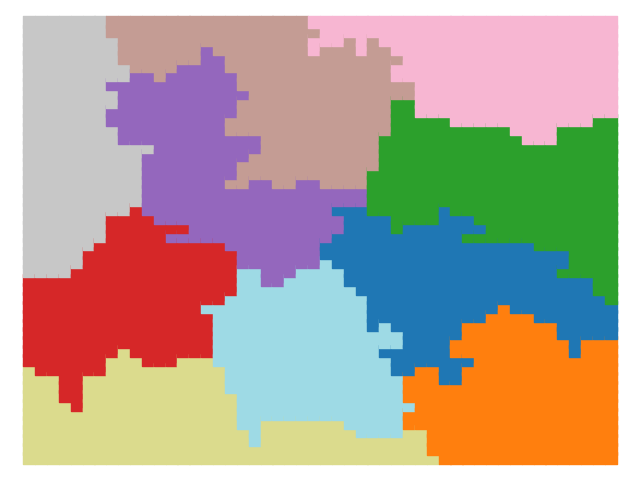}} 
\subfloat{\includegraphics[height=1.25in]{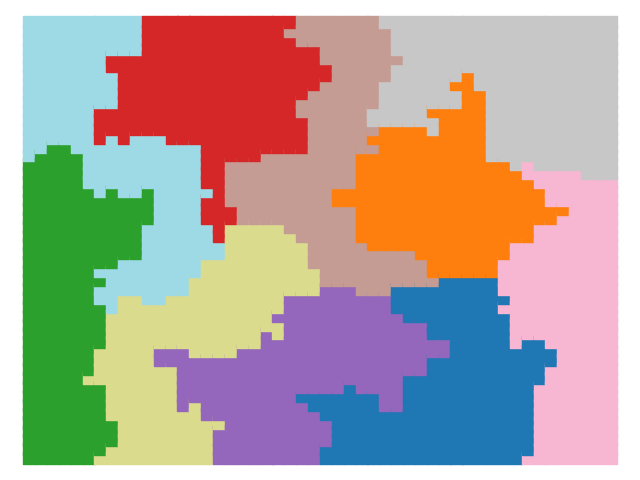}} \subfloat{\includegraphics[height=1.25in]{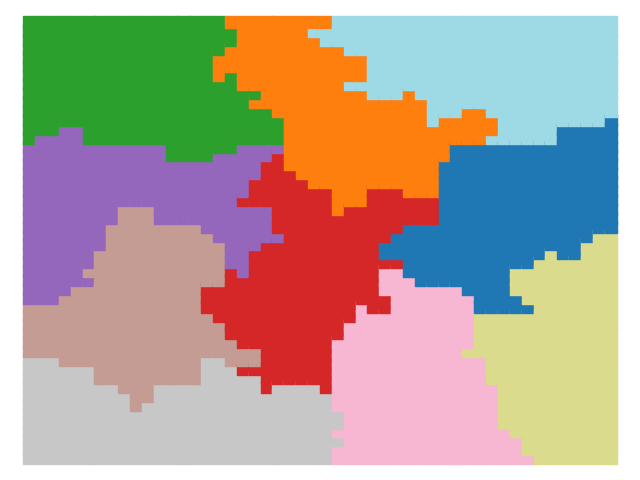}} \subfloat{\includegraphics[height=1.25in]{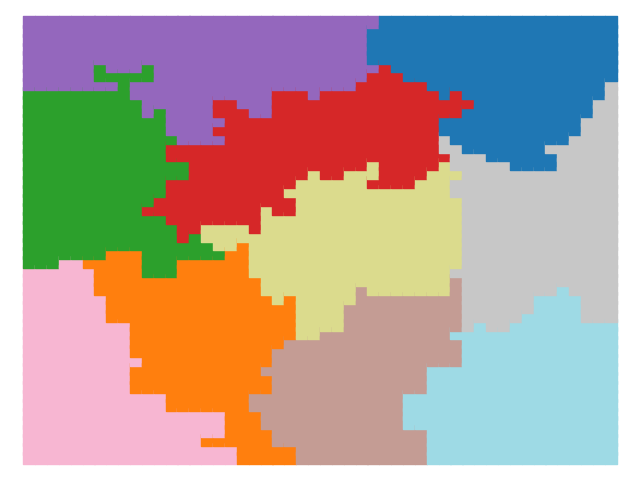}}
\caption{The \ReCom proposal tends to produce compact or geometrically tame districts, with favorable isoperimetric ratios. Each of these plans was selected after 100 \ReCom steps starting from the same vertical-stripes partition. Unlike the \Flip samples, these partitions have relatively short boundaries in addition to displaying low correlation with the initial state.}
    \label{fig:treeexample}
\end{figure}

The reason that the uniform distribution is so dominated by non-compact districts is a simple matter of numbers:
there are far more chaotic than regular partitions.  As an illustration, consider bipartitioning an $n\times n$ square grid 
into pieces of nearly the same number of nodes.  If the budget of edges you are allowed to cut is 
roughly $n$, there are a polynomial 
number of ways to bipartition, but the number grows exponentially as you relax the limitation on the boundary size.  
This exponential growth
also explains why the imposition of any strict limit on boundary length will  leave almost 
everything at or near the limit.

Spanning trees are a useful mechanism to produce contiguous partitions, since the deletion of any single edge from a tree leaves two connected subgraphs.   Furthermore, the tendency of the spanning tree process will be to produce
districts without skinny necks or tentacles.  To see this, consider the $k=2$ case first.  The number of ways for the spanning tree step to produce a 
bipartition of a graph $G$ into subgraphs $H_1$ and $H_2$ is the number of spanning trees of $H_1$ times the number
of spanning trees of $H_2$ times the number of edges between $H_1$ and $H_2$ that exist in $G$.\footnote{The idea that one can cut spanning trees to create partitions, 
and that the resulting distribution will have factors proportional to the number of trees in a block, is a very natural one and appears for instance in the ArXiv note
\url{https://arxiv.org/pdf/1808.00050.pdf}.}

Now we consider why more compact districts are up-weighted by this process.
Suppose $G$ is a graph on $N$ nodes appearing as a connected subgraph of a square lattice.  
Kirchhoff's remarkable 
counting formula for spanning trees tells us that the precise number of spanning trees 
of any graph on $N$ nodes is $\det(\Delta')$, where $\Delta'$ is any $(N-1)\times (N-1)$ 
minor of the combinatorial Laplacian $\Delta$ of $G$.
For instance, for an $n\times n$ grid, the number of spanning trees is asymptotic to $C^{n^2}=C^N$, where $C$ is a constant
whose value is roughly $3.21$ \cite{Temperley}.  There are difficult mathematical theorems that suggest that 
squares have more spanning trees than any other subgraphs of grids with the same number of nodes
\cite{Kenyon,Karlsson}.  But this means that if a district has a simple ``neck" or  ``tentacle" with just two or three nodes,
it could reduce the number of possible spanning trees by a factor of $C^2$ or $C^3$,  making the district ten or thirty
times less likely to be selected by a spanning tree process.  The long snaky districts that are observed in the \Flip ensembles
are nearly trees themselves, and are therefore dramatically down-weighted by \ReCom because they admit 
far fewer spanning trees than their plumper cousins.  
For example, the initial partition of the $50\times 50$ grid in Figure~\ref{fig:flipschematic} has roughly $10^{1210}$ spanning
trees that project to it while the final partition has roughly $10^{282}$.  That means that the tame partition is over 
$10^{900}$ times more likely to be selected by a spanning tree \ReCom step than the snaky partition, while 
uniform \Flip weights them exactly the same.%
\footnote{A similar argument should be applicable to $k>2$ districts since the basic \ReCom move handles them two at a time.}

\subsection{Complexity and mixing}\label{sec:RP=NP}

Flip distributions and uniform distributions have another marked disadvantage for sampling:  computational
intractability.  In the study of computational complexity, $P \subseteq RP \subseteq NP$ are complexity classes 
(polynomial time, randomized polynomial time, and nondeterministic polynomial time) and it is widely believed that $P = RP$, and $RP \not = NP$ . 
Recent theoretical work of DeFord--Najt--Solomon \cite{lorenzo}
shows that flip and uniform flip procedures mix exponentially slowly on several families of graphs, including
planar triangulations of bounded degree.
The authors also show that sampling proportionally to $x^{|\partial P|}$ for any $0<x\le 1$ is intractable, in the sense that an efficient solution would imply $RP=NP$.  Note that the $x=1$ case covers uniform sampling. 
This complexity analysis implies that methods targeting the uniform distribution and natural variants weighted to favor shorter boundary lengths are likely to  have complexity obstacles to overcome, particularly with respect to worst-case scenarios, raising the need for reassurance of the quality of sampling.
Our experiments in \S\ref{sec:experiments} highlight some of these challenges in a practical setting by showing that 
\Flip ensembles continue to give unstable results---with respect to starting point, run length, and summary statistics---at 
lengths in the many millions.   Practitioners must opt for fast implementations and very large subsampling time; even then, the flip approach requires dozens of tuning decisions, which undermines any sense in which  the associated stationary distribution 
is canonical.

Unlike \Flip, the \ReCom chain is designed so
that each step  completely destroys the boundary between two districts, in the sense that the previous pairwise boundary has no impact on the next step. As there are at most $\binom{k}{2}$ boundaries 
in a given $k$-partition,  this observation suggests that we can lose {\em most} memory 
of our starting point in a number of steps that is polynomial in $k$ and does not depend on $n$ at all.  
The size of the full state space of balanced $k$-partitions of an $n\times n$ grid is easily seen to be larger than exponential in $n$.
Based on a mixture of experiments and theoretical exploration, we conjecture that the full \ReCom diameter of the state space---the most steps that might be required to connect any two partitions---is sublinear (in fact, logarithmic) in $n$.  
We further conjecture that \ReCom is rapidly mixing (in the technical sense) on this family of examples,
with mixing time at worst polynomial in $n$.
(By contrast, we expect that the \Flip diameter of the state space, and its mixing time, grow exponentially or faster
in $n$.)

The Markov chain literature has examples of processes on grids with constant scaling behavior, such
as the Square Lattice Shuffle \cite{haastad2006square}.  
That chain has arrangements of $n^2$ different objects in an $n\times n$ grid as its set of states; a move consists of randomly
permuting the elements of each row, then of each column---or just one of those, then transposing.
Its mixing time is constant, i.e., independent of $n$. 
Chains with logarithmic mixing time are common in statistical mechanics: a typical fast-mixing model, like the discrete hard-core model at high temperature, mixes in time $n \log n$  with local moves (because it essentially reduces to the classic coupon collector problem), but just $\log n$ with global moves.  The global nature of \ReCom moves leaves open the possibility
of this level of efficiency.

Our experiments below support the intuition that the time needed for effective sampling has moderate growth; tens of thousands
of recombination steps  give stable results on practical-scale problems whether we work with the roughly
9000 precincts of Pennsylvania or the roughly 100,000 census blocks in our Virginia experiments.  Note that these observations do not contradict the theoretical obstructions in \cite{lorenzo}, since \ReCom is not designed to target the uniform distribution or 
any other distribution known to be intractable. 
While \ReCom is decidedly nonuniform, the arguments in \S\ref{sec:compactnessdesign} indicate that this nonuniformity is desirable, as the chain preferentially samples from plans that comport with traditional districting principles.

 \section{Experimental comparison}
\label{sec:experiments}

\begin{figure}[!h]
    \centering
    \subfloat[County]{\includegraphics[height=1.23in,width=1.5in]{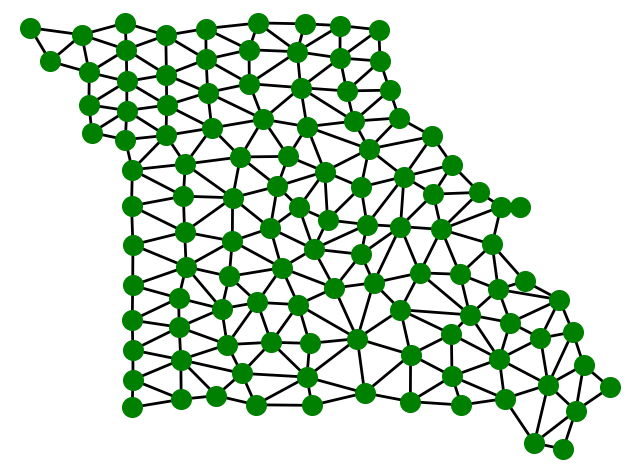}}\quad
        \subfloat[County Subunit
        ]{\includegraphics[height=1.23in,width=1.5in]{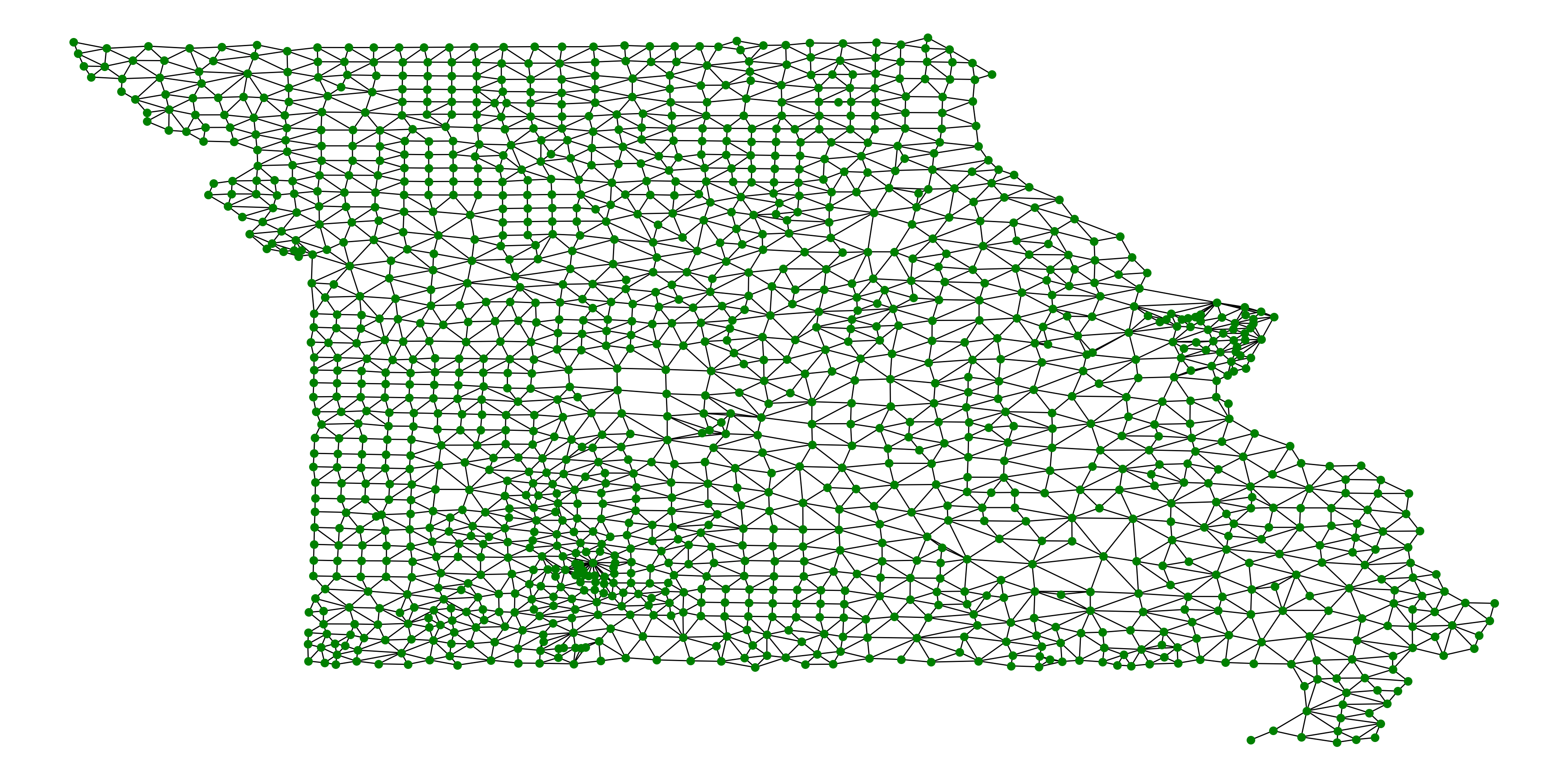}}\quad
    \subfloat[Census Tract
    ]{\includegraphics[height=1.23in,width=1.5in]{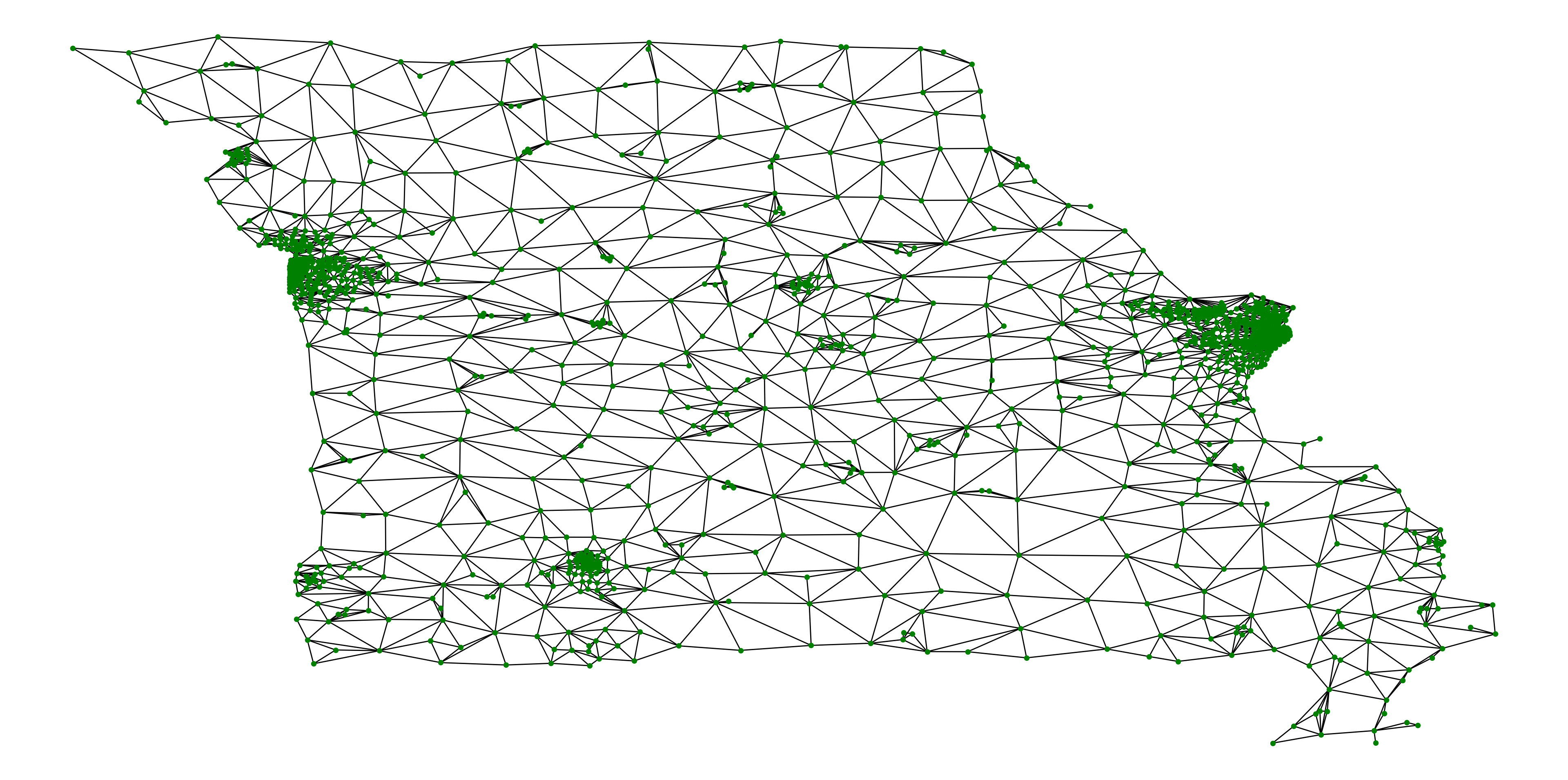}}\quad
    \subfloat[Census Block
    ]{\includegraphics[height=1.23in,width=1.5in]{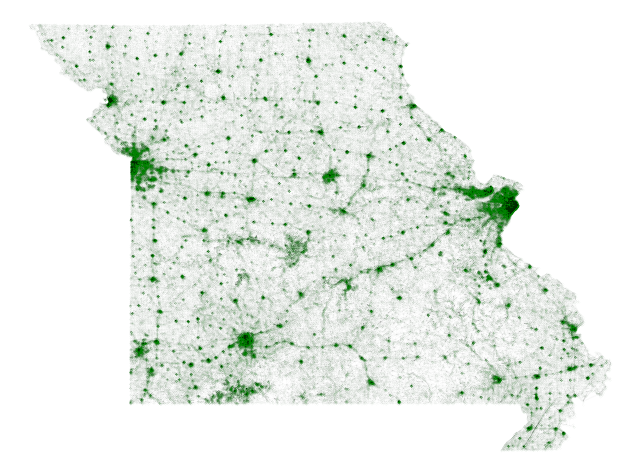}}

    \caption{Four dual graphs for Missouri at different levels of geography in the Census hierarchy. The graphs have 115, 1,395, 1,393, and 343,565 nodes respectively. }
    \label{fig:dgsquares}
\end{figure}

In this section, we will run experiments on the standard toy examples for graph problems, $n\times n$ grids, as well as on empirical dual graphs generated from census data.  The real-world graphs  can be large but they share key properties with lattices---they tend to admit planar embeddings, with most faces triangles or squares.  Figure~\ref{fig:dgsquares} shows the state of Missouri at four different levels of census geography, providing good examples of the characteristic structures we see in our applications. 

All of our experiments were carried out using the {\sf GerryChain} software \cite{gerrychain}, with additional 
source code available
for inspection \cite{replication}. The state geographic and demographic data was obtained from the census TigerLine geography program accessed through NHGIS \cite{nhgis}.

\subsection{Sampling distributions, with and without tight constraints}
\begin{figure}
    \centering
\subfloat{\includegraphics[height=1in]{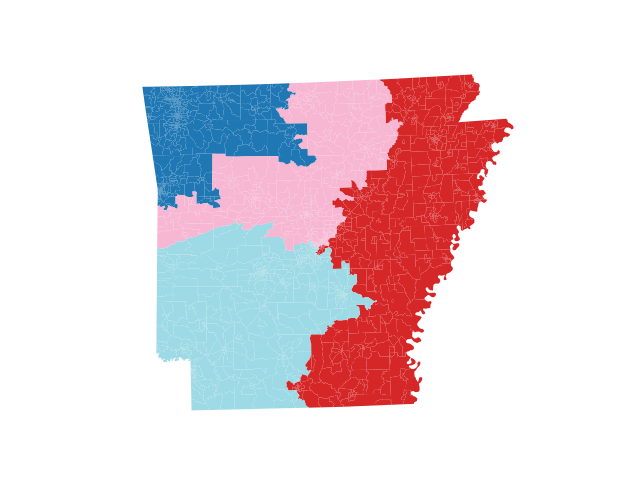}}
\subfloat{\includegraphics[height=1in]{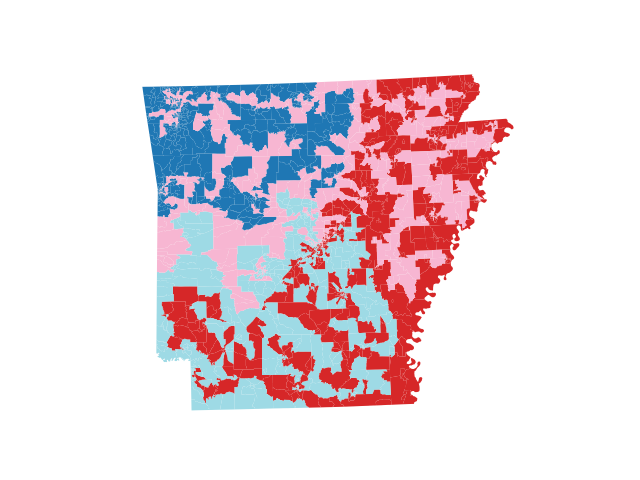}}
\subfloat{\includegraphics[height=1in]{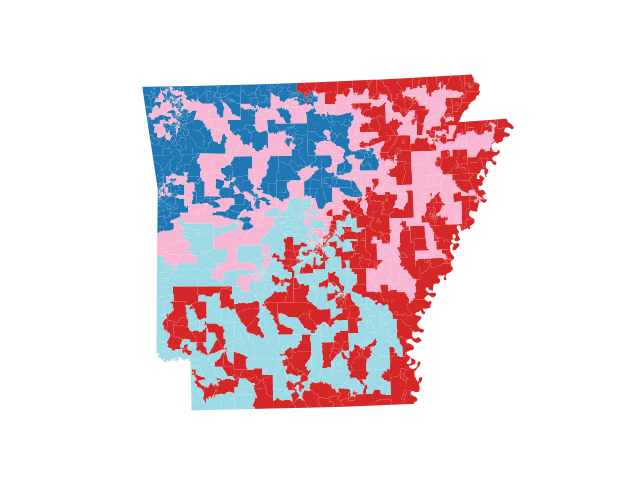}}
\subfloat{\includegraphics[height=1in]{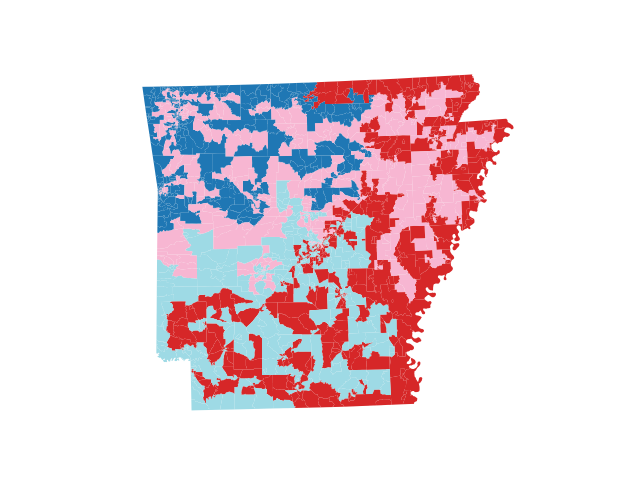}}
\subfloat{\includegraphics[height=1in]{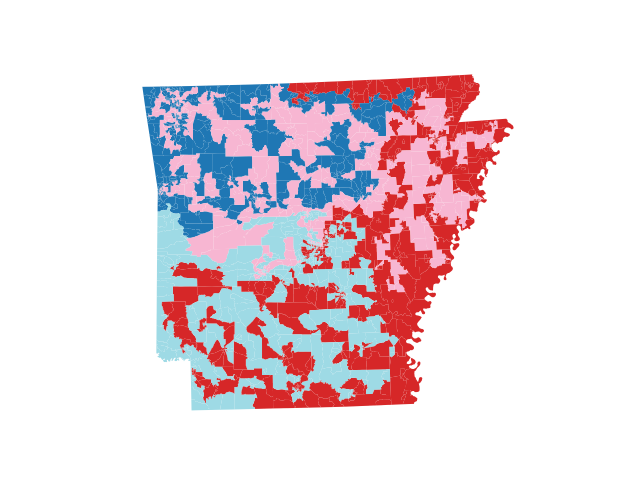}}

{\small Run 1: 100K \Flip steps, shown every 25K, no compactness constraint}

\subfloat{\includegraphics[height=1in]{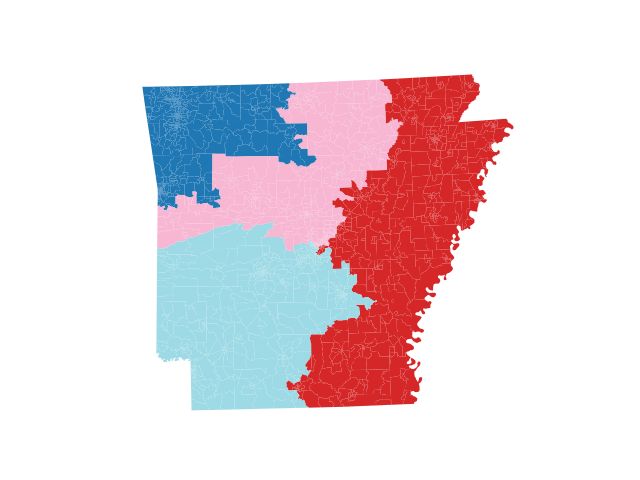}}
\subfloat{\includegraphics[height=1in]{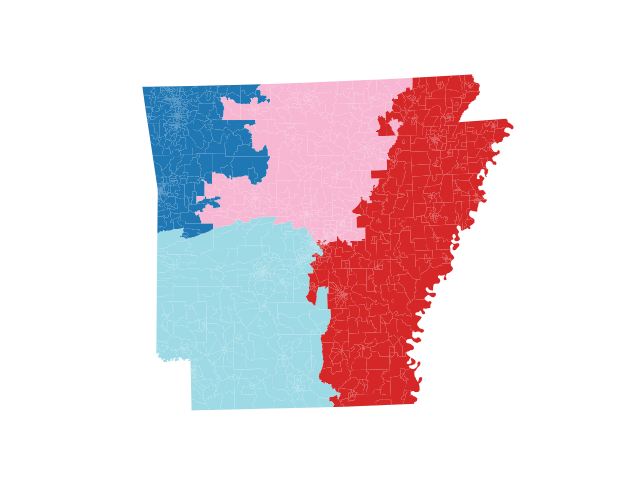}}
\subfloat{\includegraphics[height=1in]{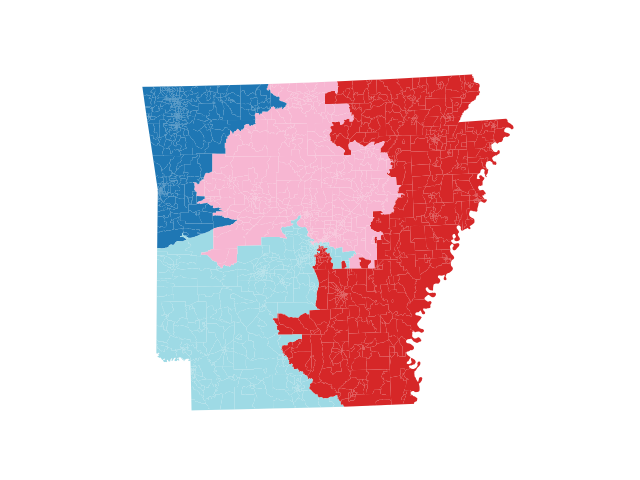}}
\subfloat{\includegraphics[height=1in]{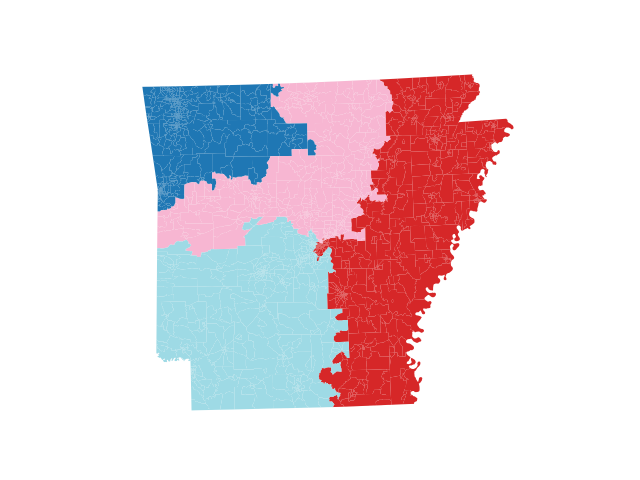}}
\subfloat{\includegraphics[height=1in]{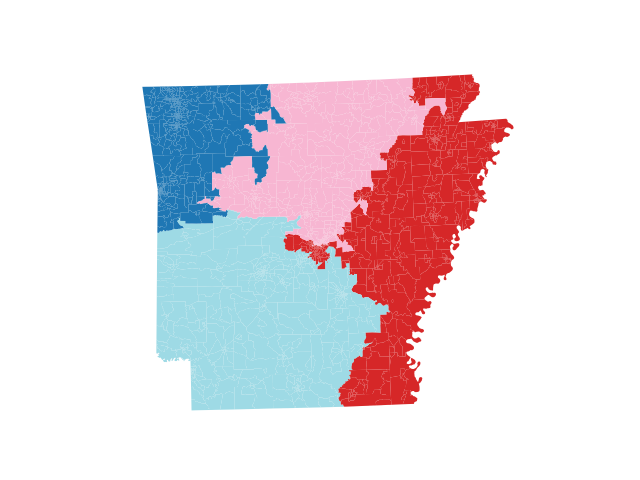}}

{\small Run 2: 100K \Flip steps, shown every 25K, limited to 5\% total cut edges}

 \subfloat{\includegraphics[height=1.2in]{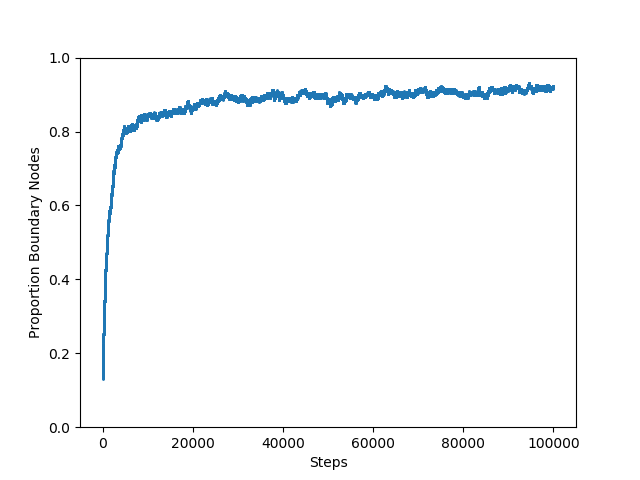}}
 \subfloat{\includegraphics[height=1.2in]{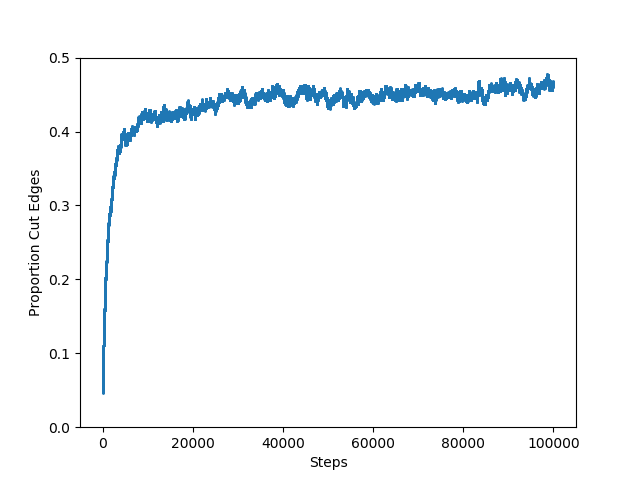}}
 \subfloat{\includegraphics[height=1.2in]{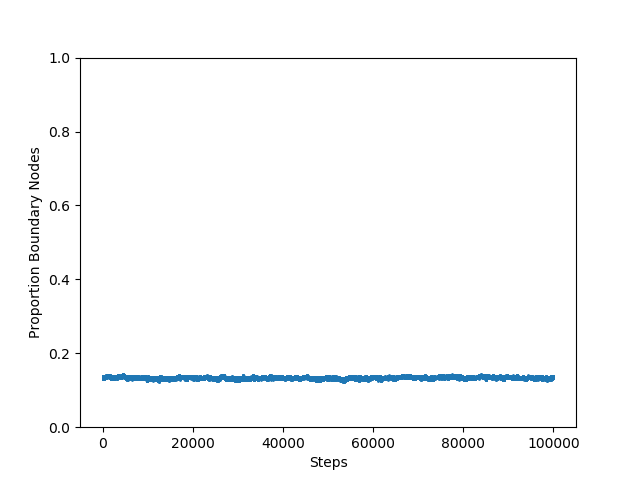}}
 \subfloat{\includegraphics[height=1.2in]{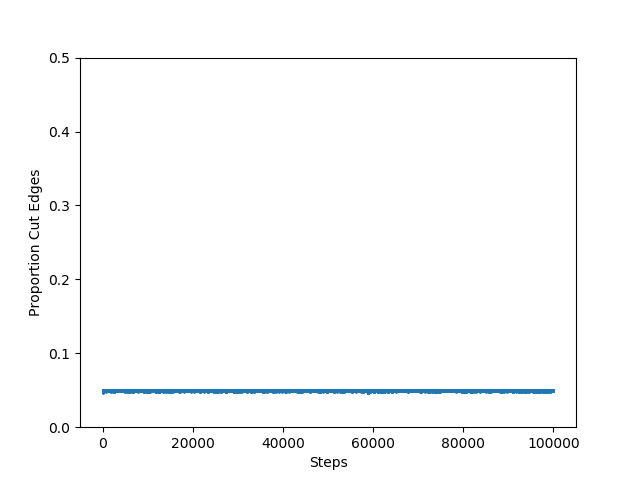}}
 
 {\small Run 1 boundary statistics} \hspace{1.5in} {\small Run 2 boundary statistics}

    \subfloat{\includegraphics[height=1in]{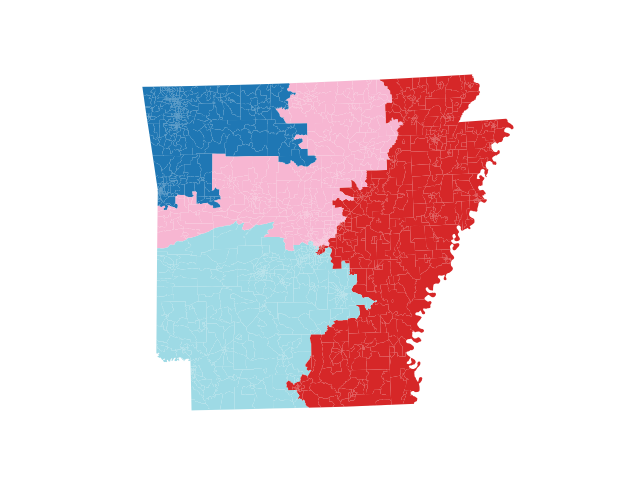}}
\subfloat{\includegraphics[height=1in]{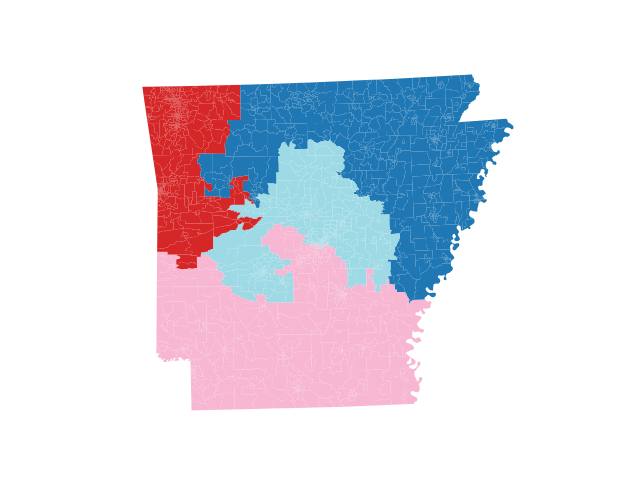}}
\subfloat{\includegraphics[height=1in]{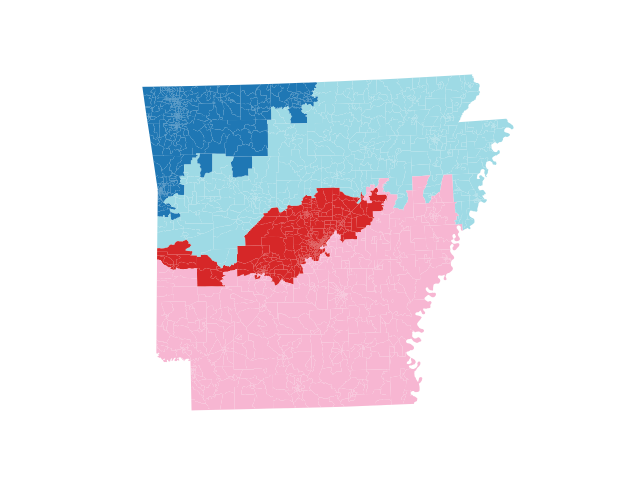}}
\subfloat{\includegraphics[height=1in]{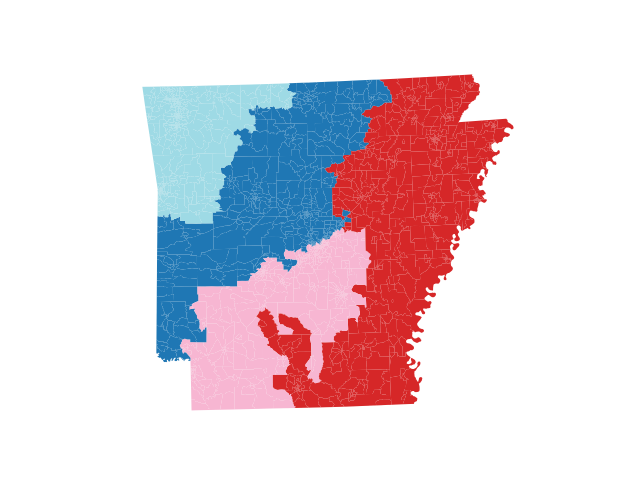}}
\subfloat{\includegraphics[height=1in]{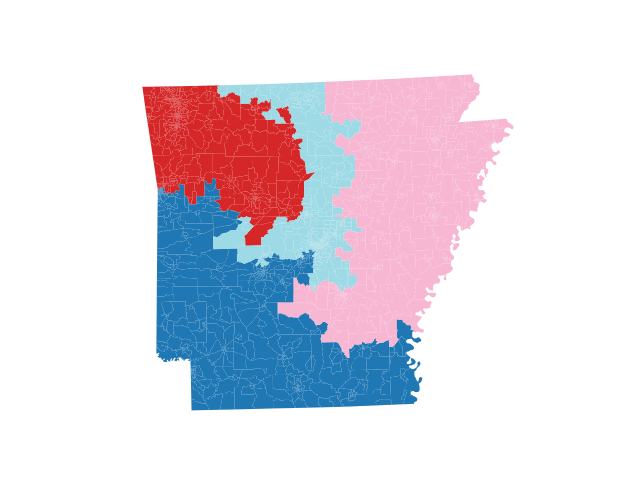}}

\small{Run 3: 10K \ReCom steps, shown every 2500, no compactness constraint}

\subfloat{\includegraphics[height=1in]{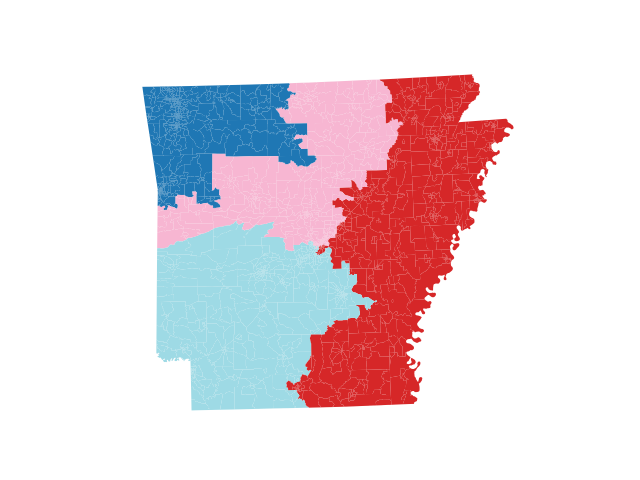}}
\subfloat{\includegraphics[height=1in]{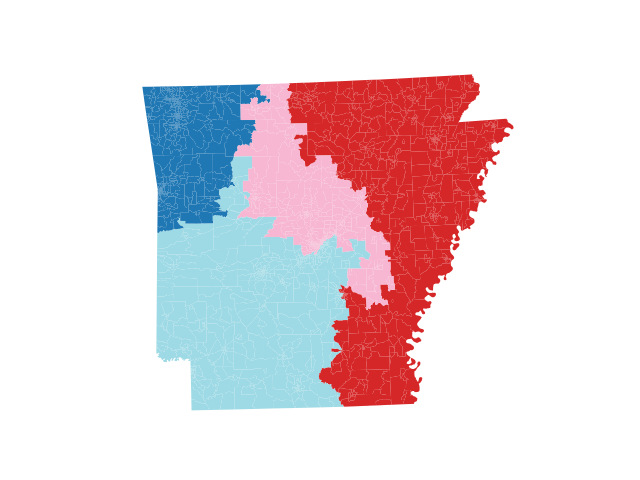}}
\subfloat{\includegraphics[height=1in]{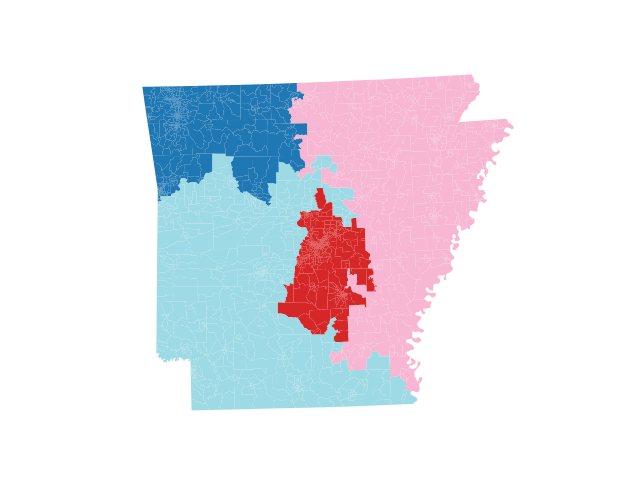}}
\subfloat{\includegraphics[height=1in]{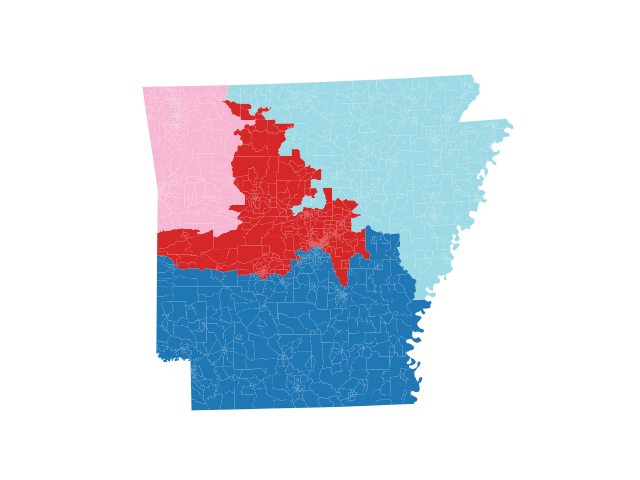}}
\subfloat{\includegraphics[height=1in]{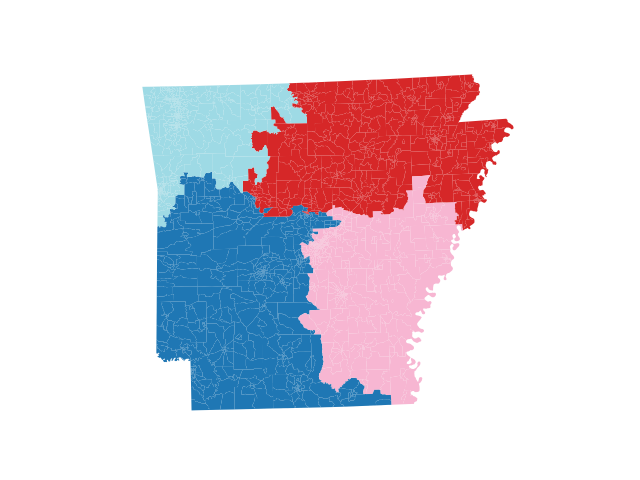}}

\small{Run 4: 10K \ReCom steps, shown every 2500, limited to  5\% total cut edges}

 \subfloat{\includegraphics[height=1.2in]{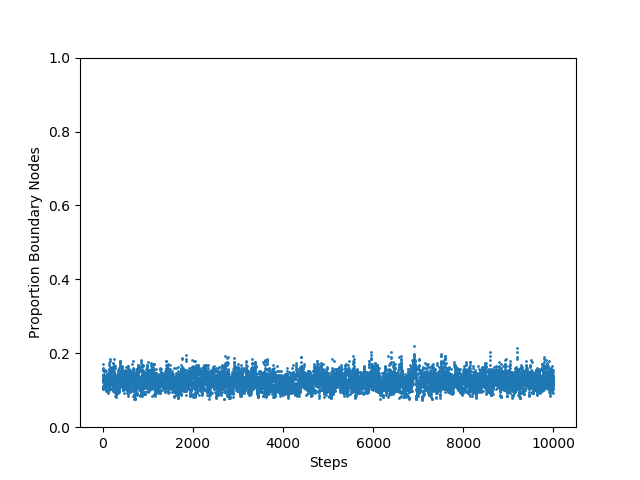}}
 \subfloat{\includegraphics[height=1.2in]{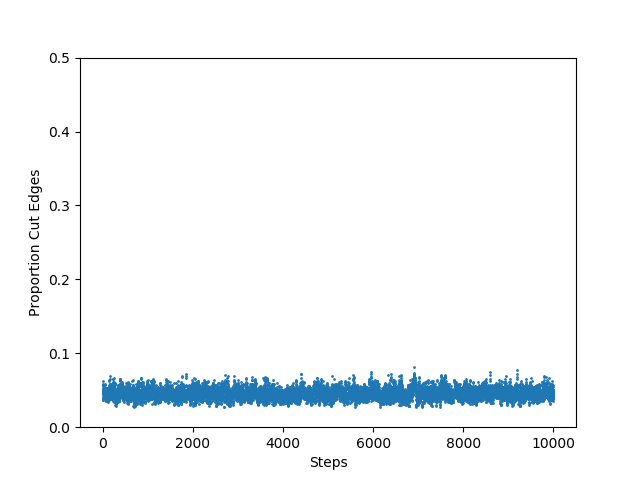}}
 \subfloat{\includegraphics[height=1.2in]{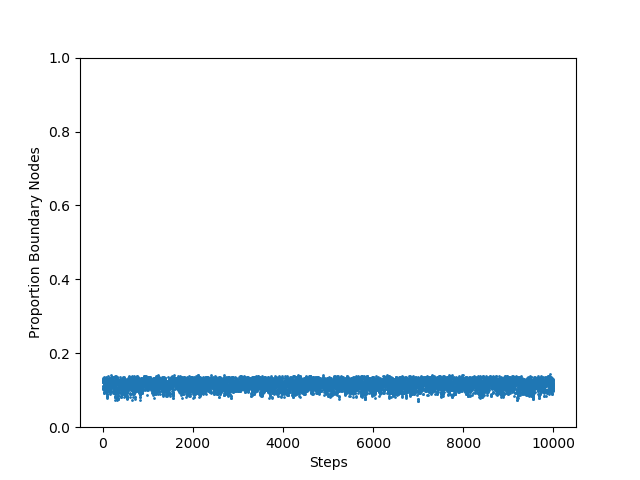}}
 \subfloat{\includegraphics[height=1.2in]{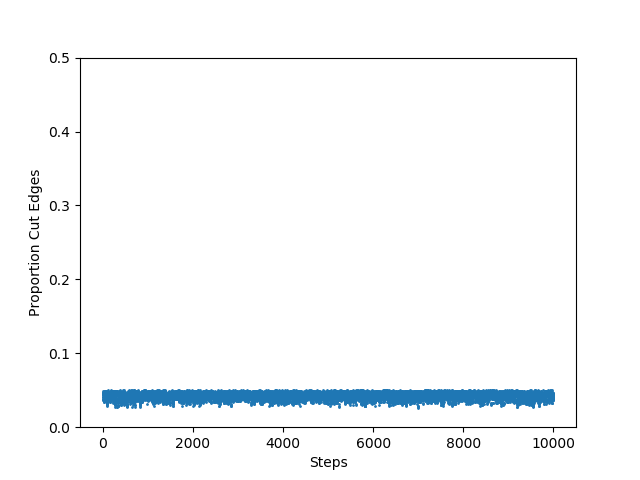}}
 
  {\small Run 3 boundary statistics} \hspace{1.5in} {\small Run 4 boundary statistics}
 
 \caption{Arkansas block groups partitioned in to 4 districts, with population deviation limited to 
 5\% from ideal.
Imposing a compactness constraint makes the \Flip chain unable to move very far.} 
    \label{fig:AR_IA}
\end{figure}

We begin by noting that the tendency of \Flip chains to draw non-compact plans is not limited to grid graphs
but occurs on geographic dual graphs just as clearly.  The first run in Figure~\ref{fig:AR_IA} shows that 
Arkansas's block groups admit the same behavior, with upwards of 90\% of nodes on the boundary of a district,
and roughly 45\% of edges cut, for essentially the entire length of the run.  
The initial plan has under 20\% boundary nodes, and around 5\% of edges cut;
the basic recombination chain (Run 3) stays right in range of those statistics.

Using thresholds or constraints to ensure that the \Flip proposals remain reasonably
criteria-compliant requires a major tradeoff. While this enforces validity, it is difficult for \Flip Markov chains to generate substantively distinct partitions under tight constraints. Instead the chain
can easily flip the same set of boundary nodes back and forth and remain in a small neighborhood around the initial plan. See the second run in Figure \ref{fig:AR_IA} for an example.  Sometimes, this is because an overly tight constraint disconnects the state space
entirely and leaves the chain exploring a small connected component.\footnote{An example of this behavior was presented in 
\cite[Fig 2]{SPP1}, though its significance was misinterpreted by the authors with respect to the test in \cite{pegden}.}
Recombination responds better to sharp constraints, and \ReCom chains do not tend to run at the limit values
when constrained.  The interactions between various choices of constraints and priorities are so far vastly under-explored.  In \S\ref{sec:metropolis}, we will consider the use 
of preferentially weighting steps rather than constraining the chains.

\subsection{Projection to summary statistics}

The space of districting plans is wildly complicated and high-dimensional.  For the redistricting application,
we are seeking  a way to understand the measurable properties of plans that have political or legal relevance, 
such as their partisan and racial statistics; this amounts 
to projection to a much lower-dimensional space.  
\begin{figure}[!h]
    \centering  
\begin{tikzpicture}    
\node at (0,1.5) {\includegraphics[height=3.3cm,width=4.3cm]{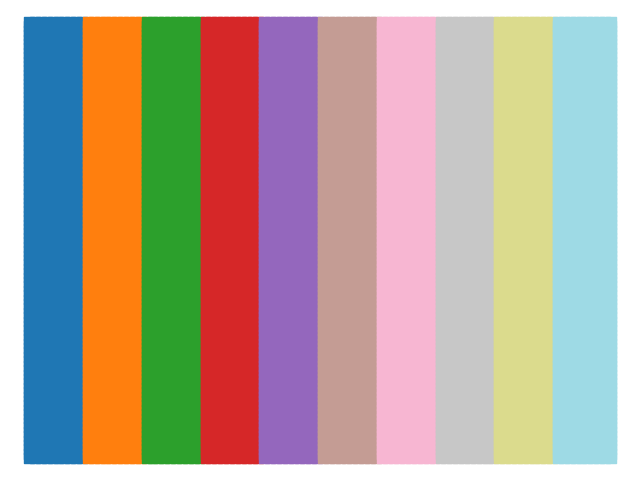}};
\draw (3,0) rectangle (7,3); 
\draw (8,0) rectangle node {$A'$} (9.6,3);
\draw (9.6,0) rectangle node {$B'$} (12,3);
\draw (8,0) rectangle (12,3); 
\draw (3,0) rectangle node {$B$} (7,1.8);
\draw (3,1.8) rectangle node {$A$} (7,3);

\node at (-1,-2) {\includegraphics[height=1.25in]{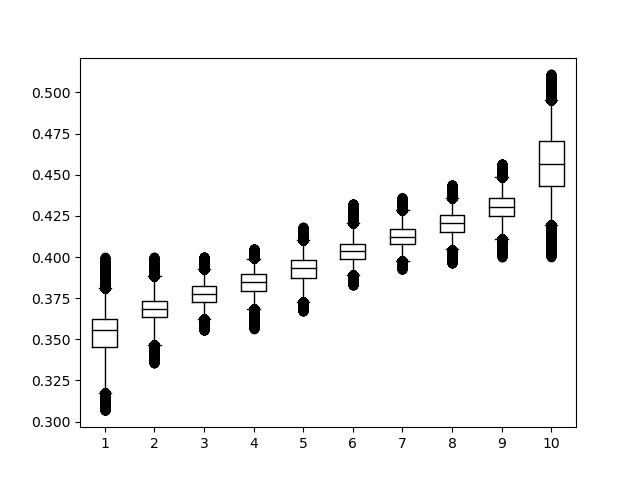}};   
\node at (-1,-4) {\Flip, $A$ share};
\node at (3,-2) {\includegraphics[height=1.25in]{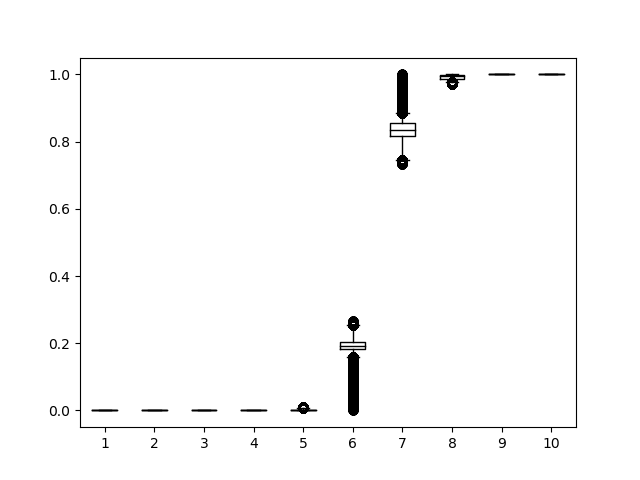}};
\node at (3,-4) {\Flip, $A'$ share};
\node at (7,-2) {\includegraphics[height=1.25in]{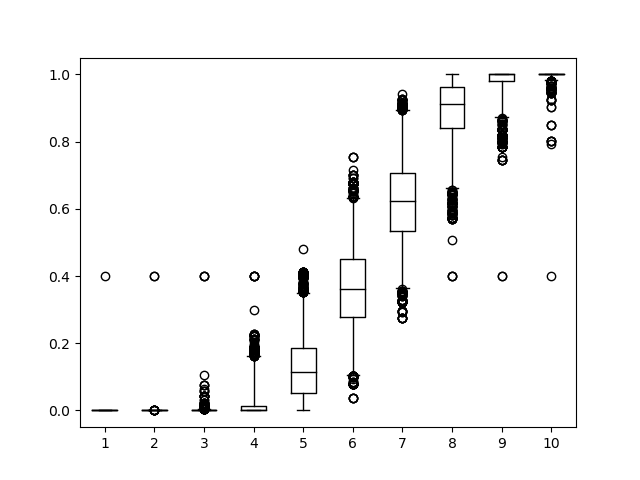}};   
\node at (7,-4) {\ReCom, $A$ share};
\node at (11,-2) {\includegraphics[height=1.25in]{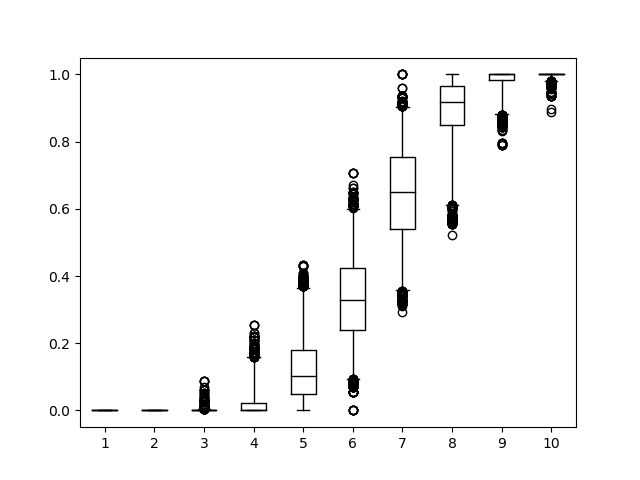}};
\node at (11,-4) {\ReCom, $A'$ share};

\node at (5,-7) {    \begin{tabular}{|c|c|c|c|c|}
    \hline
   {}& \multicolumn{2}{|c|}{\Flip}&\multicolumn{2}{|c|}{\ReCom}\\
    \hline
    \hline
    \# Seats & $A$&$A'$    &$A$&$A'$  \\
    \hline
     0    &995,158 &0 &1 &0 \\
         \hline
     1   &4,842 &0 &1 & 0\\
    \hline
     2    & 0&0 &1 &0 \\
    \hline
     3   &0 &0 &1,652 &1,574 \\
    \hline
     4   &0 &1,000,000 & 6,993&7,561 \\
    \hline
     5   & 0&0 &1,352 &865 \\
    \hline
     $\ge$ 6   & 0&0&0 &0 \\
\hline
    \end{tabular}};
\end{tikzpicture}    
    
\caption{Boxplots and a table of push-forward statistics for two synthetic elections on a grid, with one million \Flip steps
and 10,000 \ReCom steps. The boxplots show the proportion of the district made up of the group $A$ or $A'$ across
the ten districts of the plan.  The table records the number of districts with an $A$ or $A'$ majority for each plan.}
    \label{fig:grid_plots}
\end{figure}

\noindent Many of the metrics of interest on districting plans are formed by summing some value at each node of each district. For example,  the winner of an election is determined by summing the votes for each party in each geographic unit that is assigned to a given district, and so ``Democratic seats won" is a summary statistic that is real- (in fact integer-) valued.  
It is entirely plausible that chains which may mix slowly in the space
of partitions will converge much more quickly in their projection to some summary statistics.

To investigate this possibility, we begin with a toy example with synthetic vote data on a grid,
comparing the behavior of the \Flip and \ReCom proposals (Figure \ref{fig:grid_plots}). 
 For each Markov chain, we evaluate two vote distributions where each node is assigned to vote for a single party. In the first election, the votes for Party $A$ are placed in the top 40 rows of the $100\times 100$ grid, while in the second election, the votes for Party $A'$ are placed in the leftmost 40 columns of the grid.  We use the familiar vertical-stripes partition as our initial districting plan. The underlying vote data and initial partition are shown in the top row of Figure \ref{fig:grid_plots}.

The results confirm that in this extreme example the \Flip chain is unable to produce diverse election outcomes for either 
vote distribution.
Over 1,000,000 steps the \Flip ensemble primarily reported one seat outcome in each scenario: four seats in the first setup and 
zero seats in the second.  The \ReCom ensemble saw outcomes of three, four, or five seats, and the histograms are in qualitative agreement after only 10,000 steps.\footnote{We have carried out
longer runs in order to see when this observed obstruction to mixing in this particular experiment is overcome.  One billion steps seems to suffice on a $60\times 60$ grid, but not on a $70\times 70$.} 
The corresponding boxplots show a more detailed version of this story, highlighting the ways in which each ensemble captures the spatial distribution of voters.
 In both cases, the flip walk has trouble moving from its initial push-forward 
statistics, which causes it to return very different answers in the two scenarios.  The recombination walk takes just a few steps
to forget its initial position and then returns consonant answers for the two cases.
We note that this \Flip chain is far from mixed after a million steps, so the evidence here does not help us compare 
its stationary distribution to that of \ReCom.

\subsection{Weighting, simulated annealing, and parallel tempering}\label{sec:annealing}\label{sec:metropolis}

As we have shown above,  the \Flip proposal tends to create districts with extremely long boundaries, which does not produce a comparison ensemble that is   practical for our application.  To overcome this  issue, we could attempt to modify the proposal to  favor districting plans  with shorter boundaries.  As noted above, this is often done with a standard technique 
in MCMC called the \emph{Metropolis--Hastings} algorithm:  fix a compactness score, such as a 
 notion of boundary length $|\partial P|$, prescribe a distribution proportional to $x^{|\partial P|}$ on the state space,
 and use the Metropolis--Hastings rule to preferentially accept more compact plans.  
As discussed  above in \S\ref{sec:RP=NP}, there are computational obstructions to 
sampling proportionally to $x^{|\partial P|}$ \cite{lorenzo}. Even if we are unable to achieve a perfect sample from this distribution, however, it could be the case that this strategy generates a suitably diverse ensemble in reasonable time for our applications.

The \Flip distribution was already slow to mix, and Metropolis--Hastings adds an additional score computation and accept/reject decision at every step to determine whether to keep a sample; this typically implies that this variant runs more slowly than the unweighted proposal distribution.  To aid in getting reliable results from slow-mixing systems, it is common practice to employ another MCMC variant called \emph{simulated annealing}, which iteratively tightens the prescribed distribution toward the desired target---effectively taking larger and wilder steps initially to promote randomness, then becoming gradually more restrictive.

\begin{figure}[!h]
    \centering
    
    \subfloat{\includegraphics[width=.3\textwidth]{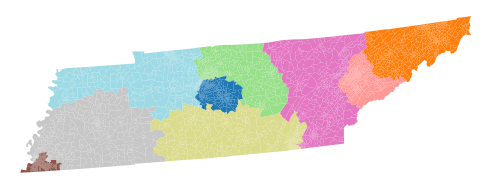}}\quad
    \subfloat{\includegraphics[width=.3\textwidth]{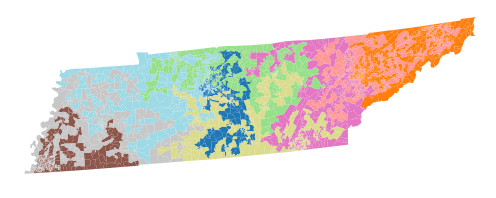}}\quad
    \subfloat{\includegraphics[width=.3\textwidth]{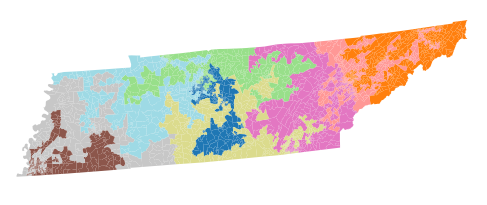}}\\
    \subfloat{\includegraphics[width=.3\textwidth]{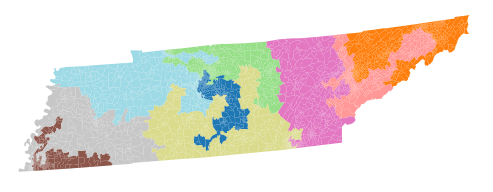}}\quad
    \subfloat{\includegraphics[width=.3\textwidth]{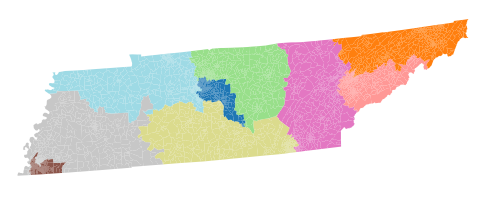}}\quad
    \subfloat{\includegraphics[width=.3\textwidth]{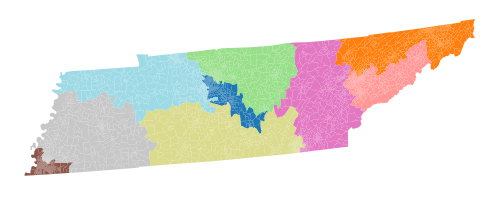}}
    
\subfloat[TN Nodes]{\includegraphics[height=1.25in]{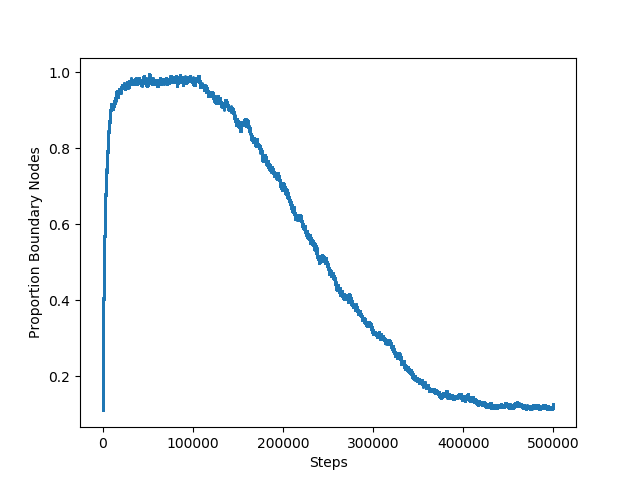}}
\subfloat[TN Edges]{\includegraphics[height=1.25in]{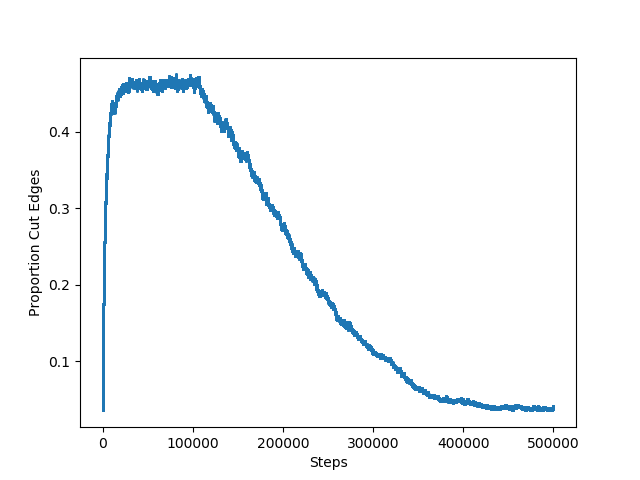}} 
\subfloat[KY Nodes]{\includegraphics[height=1.25in]{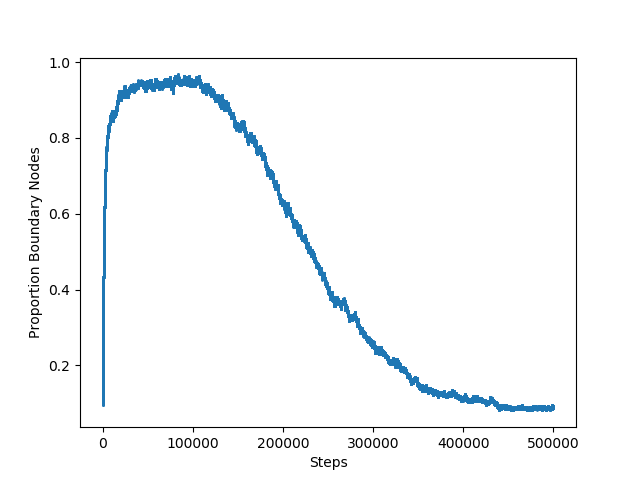}}
\subfloat[KY Edges]{\includegraphics[height=1.25in]{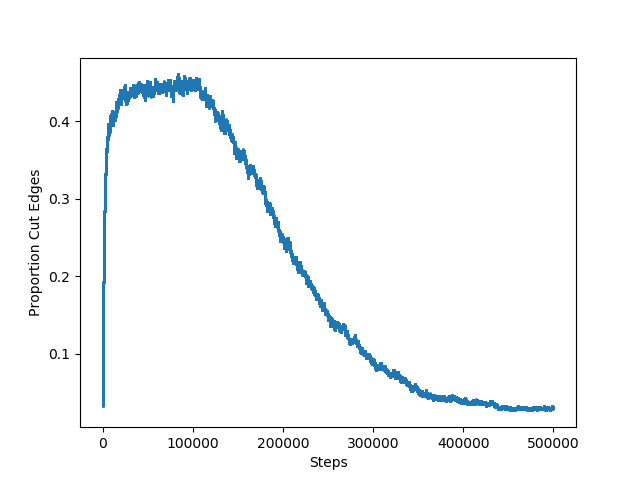}}

        \subfloat{\includegraphics[width=.3\textwidth]{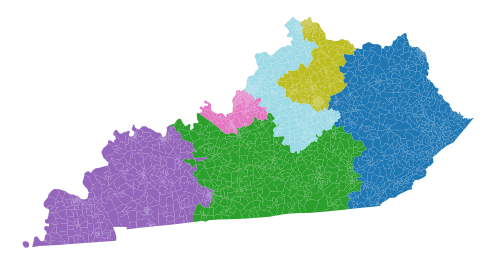}}\quad
    \subfloat{\includegraphics[width=.3\textwidth]{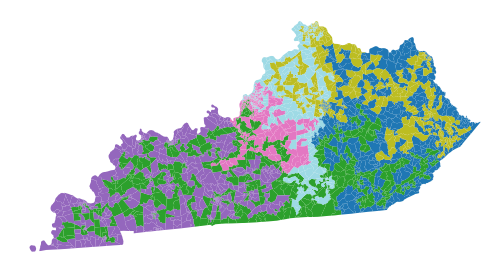}}\quad
    \subfloat{\includegraphics[width=.3\textwidth]{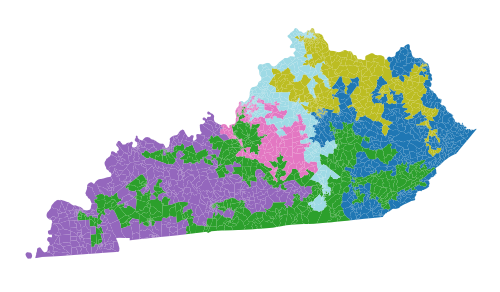}}\\
    \subfloat{\includegraphics[width=.3\textwidth]{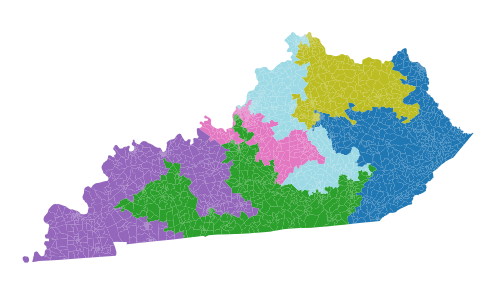}}\quad
    \subfloat{\includegraphics[width=.3\textwidth]{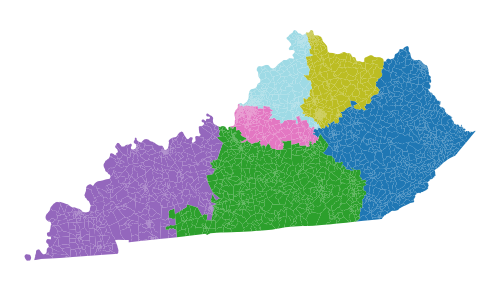}}\quad
    \subfloat{\includegraphics[width=.3\textwidth]{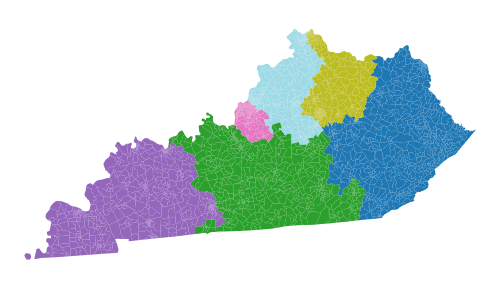}}
    \caption{Snapshots of the TN and KY annealing ensembles after each 100,000 steps.  Comparing the starting and ending states shows only slight changes to the plans as a result of the boundary segments mostly remaining fixed throughout the chain.   }
    \label{fig:anneal_plots}
\end{figure}

To test the properties of a simulated annealing run based on a Metropolis-style weighting,
we run chains to partition Tennessee and Kentucky block groups into nine  and six Congressional districts, respectively.
We run the \Flip walk for 500,000 steps beginning at a random seed drawn by the recursive tree method. 
The first 100,000 steps use an unmodified \Flip proposal; Figure \ref{fig:anneal_plots} shows that after this many steps,
the perimeter statistics are comparable to the Arkansas outputs above, with over 90\% boundary nodes and nearly 
50\% cut edges. 
This initial phase is equivalent to using an acceptance function proportional to $2^{\beta |\partial P|}$ with $\beta=0$. 
The remainder of the chain linearly interpolates $\beta$ from 0 to 3 along the steps of the run.

Figure \ref{fig:anneal_plots} shows how these Tennessee and Kentucky chains evolved. Ultimately, there is a relatively small difference between the initial and final states in both examples:  the simulated annealing has caused the random walk to return to very near its start point. This is due to the properties of the \Flip proposal. The districts grow tendrils into each other, but the boundary segments rarely change assignment. Thus, when the annealing forces the tendrils to retract, they collapse near the original districts, and this modified \Flip walk has failed to move effectively through the space of partitions. 

Other ensemble generation approaches such as \cite{Fifield_A_2018} use \emph{parallel tempering} (also known as 
{\em replica exchange}), a related technique in MCMC also aimed at accelerating its dynamics.  In this algorithm, chains are run 
in parallel from different start points at different temperatures, then the temperatures are occasionally exchanged.  Exactly the issues highlighted above apply to the individual chains in a parallel tempering run, making this strategy struggle to 
introduce meaningful new diversity.

These experiments suggest that the tendency of \Flip chains to produce fractal shapes is extremely difficult to remediate
and that direct attempts to do so end up impeding any progress of the chain through the state space.
On moderate-sized problems, this can conceivably be countered with careful tuning and extremely long runs.  
By contrast, \ReCom generates plans with relatively few cut edges (usually comparable to human-made plans) by default, and our experiments indicate that its samples are uncorrelated after far fewer steps of the chain---hundreds rather than many
billions.  
Weighted variants of \ReCom can then be tailored to meet other principles by modifying the acceptance probabilities
to favor higher or lower compactness scores, or the preservation of larger units like counties and communities of interest.
With the use of constraints and weights, one can effectively use \ReCom to impose and compare all
 of the redistricting rules and 
priorities described above \cite{Alaska,VA-report,VA-criteria}.\footnote{The weighting of a spanning tree \ReCom 
chain is not implemented with a full (reversible) Metropolis-Hastings algorithm for the same reason that the chain is not reversible
in the first place:  it is not practical to compute all of the transition probabilities from a given state in this implementation.  Nevertheless
a weighting scheme preserves the Markov property and passes the same heuristic convergence tests as before. Several teams 
are now producing Recombination-like algorithms that are reversible and still fairly efficient (references to be added when available).}

\newpage
\section{Case study:  Virginia House of Delegates}

Finally, we demonstrate the assessment of convergence diagnostics 
and the analysis  enabled by a high-quality comparator ensemble in a redistricting problem of current legal interest.   
For details, see \cite{VA-report}; we include a brief discussion with updated data here in Appendix~\ref{sec:appendix}.

The districting plan for Virginia's 100-member House of Delegates was commissioned and enacted by its 
state legislature in 2011, following the 2010 Census.  That plan was challenged in 
 complicated litigation that went before multiple federal courts before reaching the Supreme Court earlier this year,
with the ultimate finding that the plan was an unconstitutional racial gerrymander.  
The core of the courts' reasoning was that it is impermissible for the state to have constructed the districts in such a way 
that Black Voting Age Percentage (or BVAP) hit the 55\% mark in twelve of the districts.  Defending the enacted plan,
the state variously claimed that the high BVAP 
was necessary for compliance with the Voting Rights Act and that it was a natural consequence of the 
state's geography and demographics. 
The courts disagreed, finding that 55\% BVAP was
unnecessarily elevated in 11 of 12 districts, and that it caused dilution of the Black vote elsewhere in the state.

In Appendix~\ref{sec:appendix}, we present various kinds of evidence, focusing on the portion of the state
covered by the invalidated districts and their neighbors.
To assess the possibility that the BVAP is excessively elevated in the top 12 districts without algorithmic output to help,
other human-made plans can be used for comparison, even though these are limited in number and  their designers
may have had their own agendas.\footnote{Besides the original enacted 
plan, a sequence of replacement proposals introduced in the legislature, reform plans proposed by the NAACP and the
Princeton Gerrymandering Project, and finally the plan drawn by a court-appointed special master.}    
Alternately, one could make the observation that the enacted
plan's BVAP values suspiciously jump the 37-55\% BVAP range, the same range that expert reports indicate might be plausibly necessary for Black residents to elect candidates of choice.  (See Figure~\ref{fig:proposed_plans}, which shows the alternative 
proposed plans and this key BVAP range on the same plot.)  But neither of these 
adequately controls for the effects of the  actual distribution of Black population across the state geography.  
For this task, we can employ a large, diverse ensemble of alternatives made without consideration of racial statistics.

Figures~\ref{fig:compare}--\ref{fig:mms} demonstrate that for all the reasons shown in the simpler experiments above---compactness, failure of convergence in projection to racial or partisan statistics---individual \Flip chains do not produce diverse ensembles,  while \ReCom chains pass tests of quality.\footnote{The  metric used in Figure~\ref{fig:mms} 
is called the mean-median score; it is a signed measure of party advantage that is one of the leading partisan metrics
in the political science literature.}
In Figure~\ref{fig:VA-ensembles}, we apply the \ReCom outputs,  studying the full ensemble  (top plot)
and the winnowed subset of the ensemble containing only plans in which no district exceeds 60\% BVAP (bottom).
This finally allows us to address two key points with the use of an appropriate counterfactual.  
First, the BVAP pattern in the enacted plan is not explained by the human geography of Virginia.
Also, since the top 12 districts have elevated BVAP compared to the neutral plans, we can locate the costs across
the remaining districts:  it is the next four districts and even the nine after that that exhibit depressed BVAP, supporting claims of 
vote dilution.  This gives us evidence of the classic gerrymandering pattern of ``packing and cracking"---overly 
concentrated population in some districts and dispersed population in other districts that were near to critical mass.
We emphasize that ensemble analysis does not stand alone in the study of gerrymandering, but it provides a unique
ability to identify outliers against a suitable counterfactual of alternative valid plans, holding political and physical
geography constant.

%
%
%
%
%
%
%
%
%
%
%
%
%
%
%
%
%
%

\section{Discussion and Conclusion}

Ensemble-based analysis provides much-needed machinery for understanding districting plans in the context of viable
alternatives:  by assembling a diverse and representative collection of plans, 
we can  learn about the range of possible district properties along several axes, from partisan balance to shape to demographics. When a proposed plan is shown to be an extreme outlier relative to a population of alternatives, we may infer that the plan is better explained by goals and principles that were not incorporated in the model design.

Due to the extremely large space of possible districting plans for most states, we can come nowhere close to complete
enumeration of alternatives.  
For this reason, the design of an ensemble generation algorithm is a subtle task fraught with mathematical, statistical, and computational challenges.  
Comparator plans must be legally viable and pragmatically plausible to draw any power from the conclusion
that a proposed plan has very different properties.  
Moreover, to promote consistent and reliable analysis, it is valuable to connect the sampling method to a well-defined distribution over plans that not only has favorable qualitative properties but also can be sampled tractably.  This consideration leads us to study mixing times, which bolster confidence that a sample drawn using MCMC comes from the prescribed stationary distribution. 

Across a range of small and large experiments with synthetic and observed data, we find that a run assembled in several
days on a standard laptop produces \ReCom ensembles whose measurements do not vary substantially between trials,
whether re-running to vary the sample path through the state space or re-seeding at a new starting point.

Many interesting questions remain to be explored.  Here is a selection of open questions and research directions.
\paragraph*{Mathematics}
\begin{itemize}
\item Explore the mathematical properties of spanning tree bipartitioning.  For instance, what proportion of spanning trees in a grid have an edge whose complementary components have the same number of nodes?  Describe the distribution on 
2-partitions induced by trees thoroughly enough to specify a reversible version of \ReCom.
\item Describe the stationary distribution for spanning tree \ReCom.  In particular, many experiments show that 
the number of cut edges appears to be normally distributed in a \ReCom ensemble (see Fig~\ref{fig:chain_ce}(b)).  
Prove a central limit theorem for boundary length in 
\ReCom sampling of $n\times n$ grids into $k$ districts, with parameters depending on $n$ and $k$.
\item Prove rapid mixing of \ReCom for the grid case.  Even ergodicity of the chain and diameter bounds for the state
space are difficult open questions.
\item Find conditions on summary statistics that suffice for faster mixing in projection than in the full
space of plans.  Relatedly, find a class of ensemble generation techniques that would suffice to get repeatable results
in projection even if the ensembles are not similar in the space of plans.
\end{itemize}

\paragraph*{Computation}
\begin{itemize}
\item Propose other balanced bipartitioning methods to replace spanning trees, supported by fast algorithms. 
Subject these methods to similar tests of quality:  adaptability to districting principles, convergence in projection
to summary statistics independent of seed, etc.
\item Find effective parallelizations to multiple CPUs while retaining control of the sampling distribution.
\end{itemize}

\paragraph*{Modeling}
\begin{itemize}
 \item Study the stability of \ReCom summary statistics to perturbations of the underlying graph.  This  ensures that ensemble analysis is robust to some of the implementation decisions made when converting geographical data to a dual graph.
\item Identify sources of voting pattern data (e.g., recent past elections) and summary statistics (e.g., metrics in the political
science literature) that best capture the signatures of racial and partisan
gerrymandering.
\item Consider whether these analyses can be gamed:
could an adversary with knowledge of a Markov proposal create plans that are extreme in a way that is hidden, avoiding
 an outlier finding?
 \end{itemize}

\ReCom is available for use as an open-source software package, accompanied by a suite of tools to process maps and facilitate MCMC-based analysis of plans.  Beyond promoting adoption of this methodology for ensemble generation, we aim to use this release as a model for open and reproducible development of tools for redistricting.  By making code and data public, we can promote public trust in expert analysis and facilitate broader engagement among the many interested parties  in the redistricting process.

\section*{Acknowledgments}

We are grateful to the many individuals and organizations whose discussion and input informed our approach to this work.  We  thank  Sarah Cannon, Sebastian Claici,  Lorenzo Najt, Wes Pegden, Zach Schutzman, Matt Staib,  Thomas Weighill,  and  Pete Winkler for wide-ranging conversations about spanning trees, Markov chain theory, MCMC dynamics, 
and  the interpretation of ensemble results.  We are grateful to Brian Cannon for his help and encouragement
in making our Virginia analysis relevant to the practical reform effort.
The GerryChain software accompanying this paper was initiated by participants in the Voting Rights Data Institute (VRDI) 2018  at Tufts and MIT, 
and we are deeply grateful for their hard work, careful software development, and ongoing involvement.  
Max Hully and Ruth Buck were deeply involved in the data preparation and software development that made the 
experiments possible.  Finally, we acknowledge the generous support of the Prof.\ Amar G.\ Bose Research Grant 
and the Jonathan M.\ Tisch College of Civic Life.

\bibliography{refs}{}
\bibliographystyle{alpha}

\newpage
\appendix
\section{Plots for Virginia Case Study}\label{sec:appendix}

\begin{figure}[!h]
\centering
\makebox[\textwidth][c]{       \subfloat[HB5005 (Enacted)]{\includegraphics[width=2in]{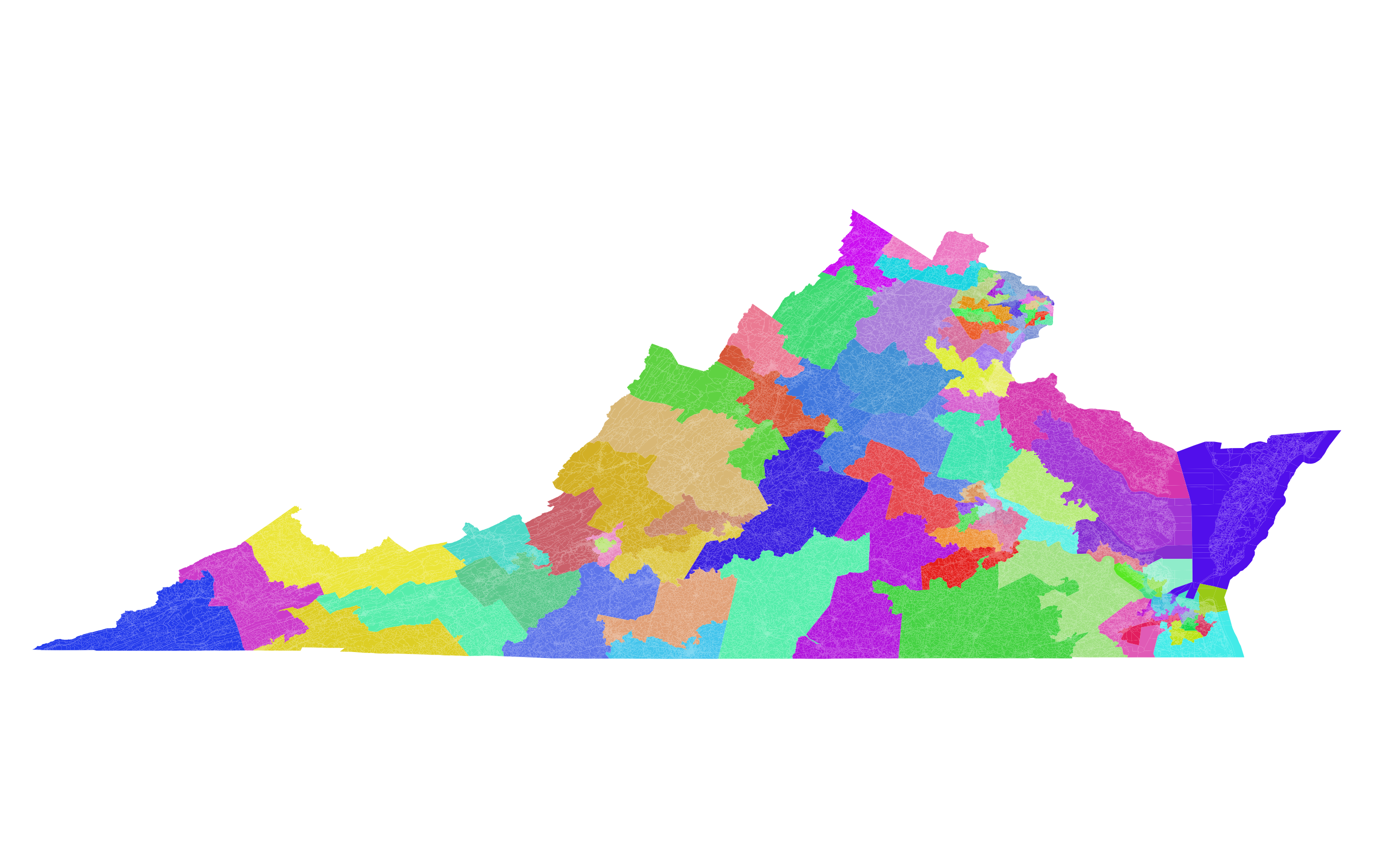}}
    \subfloat[HB7002 (Dem)]{\includegraphics[width=2in]{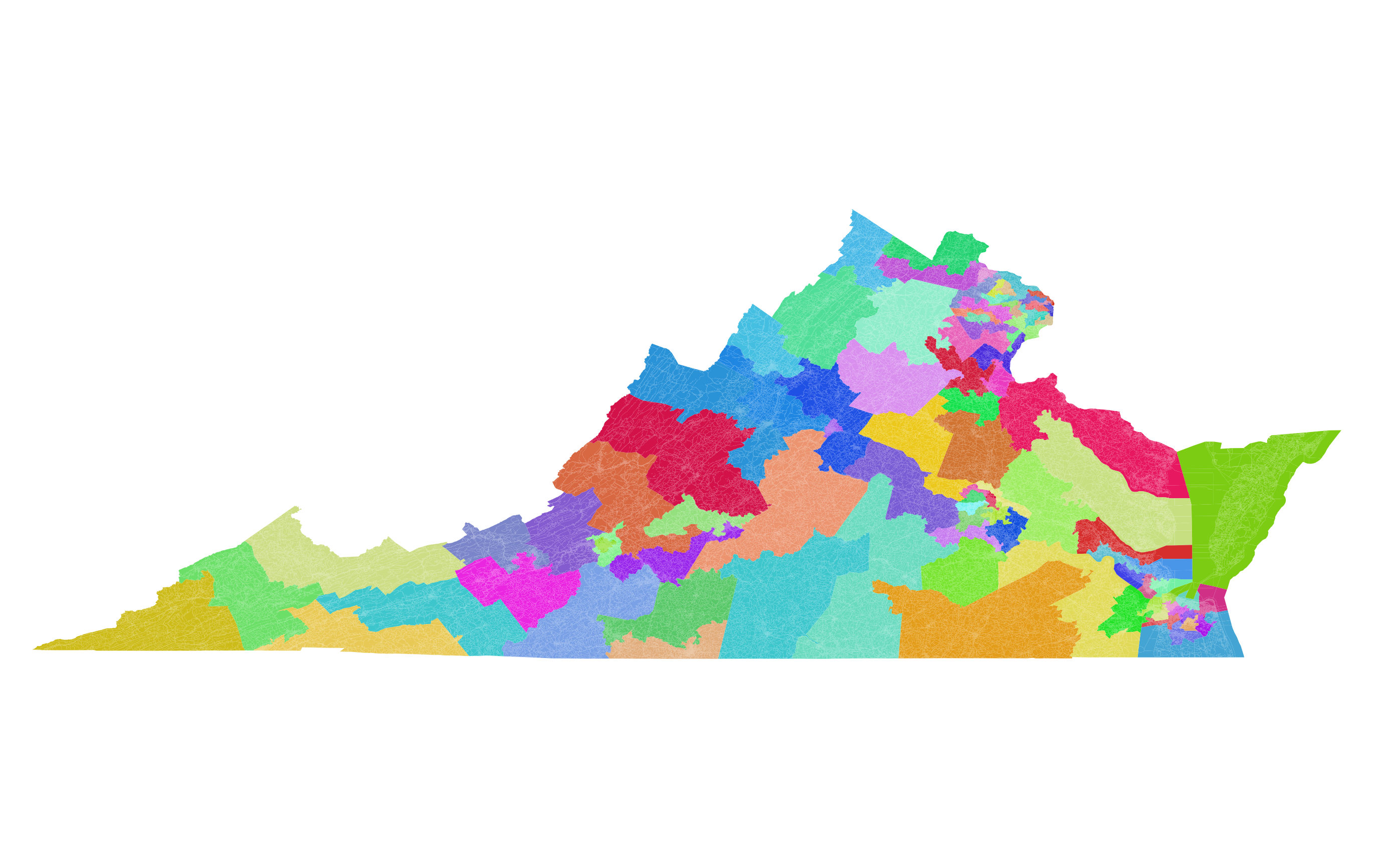}}
    \subfloat[HB7002ANS (GOP1)]{\includegraphics[width=2in]{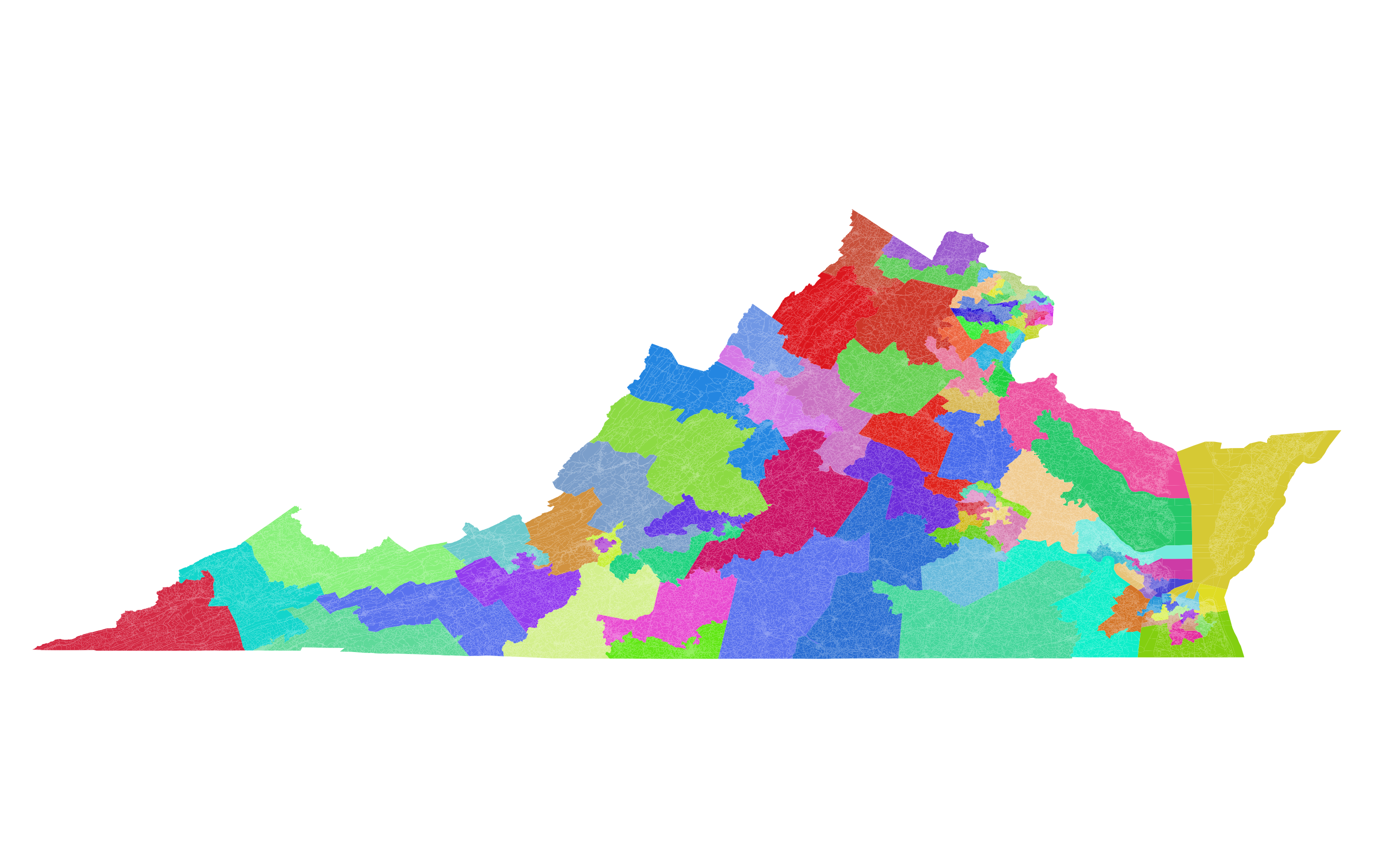}}
    \subfloat[HB7003 (GOP2)]{\includegraphics[width=2in]{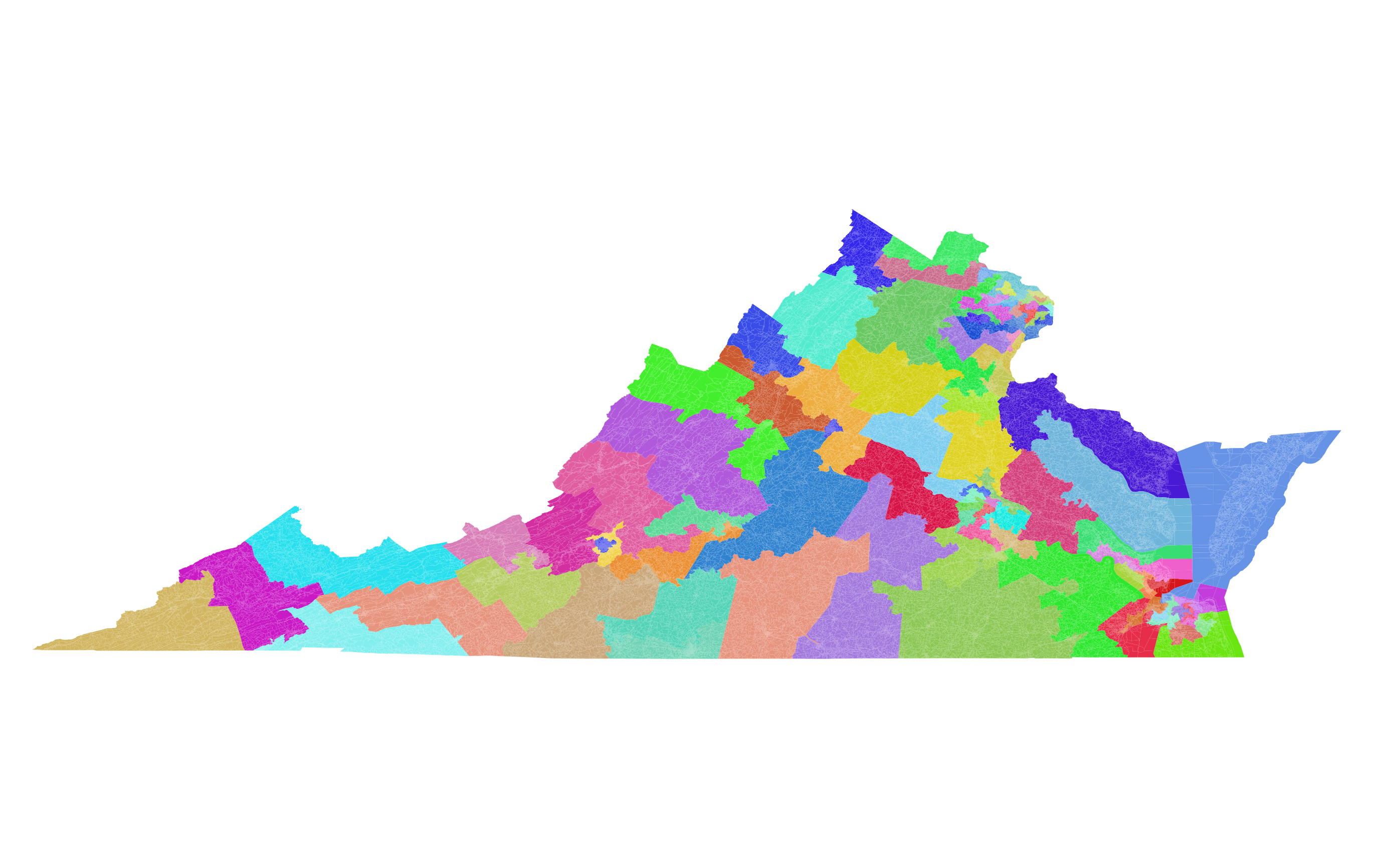}}
}\\
\makebox[\textwidth][c]{   
        \subfloat[HB7001 (GOP3)]{\includegraphics[width=2in]{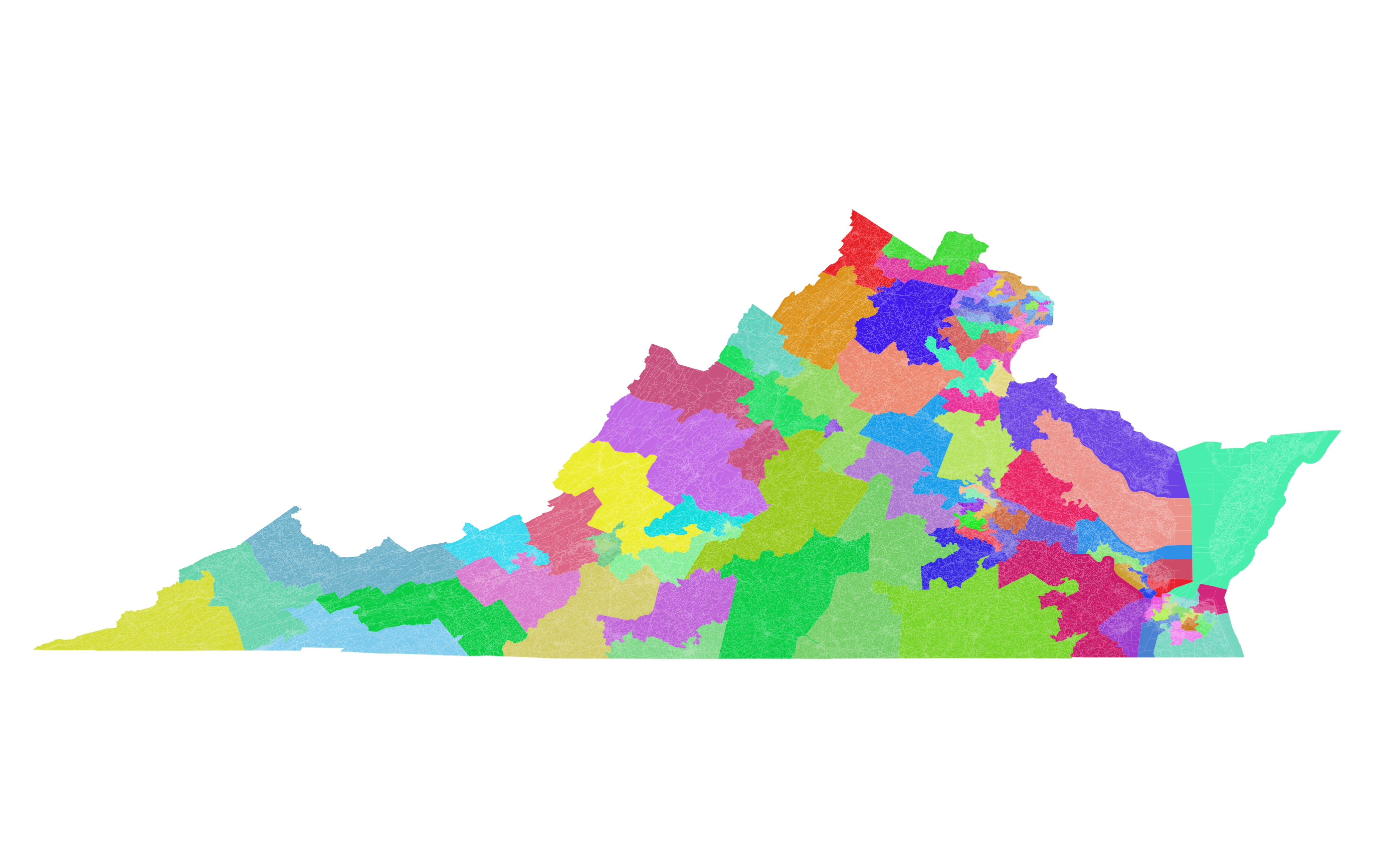}}
        \subfloat[NAACP]{\includegraphics[width=2in]{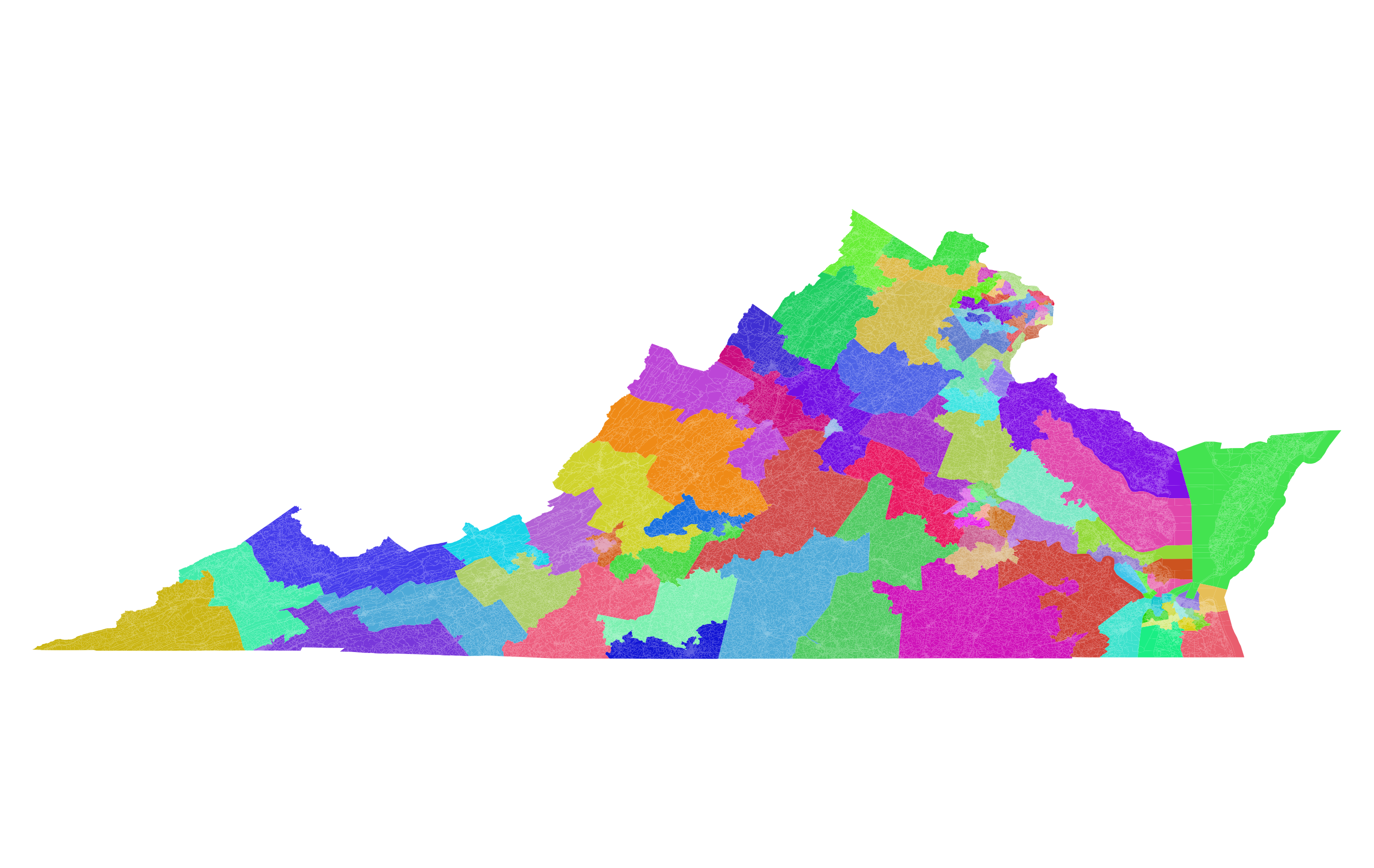}}
        \subfloat[Princeton]{\includegraphics[width=2in]{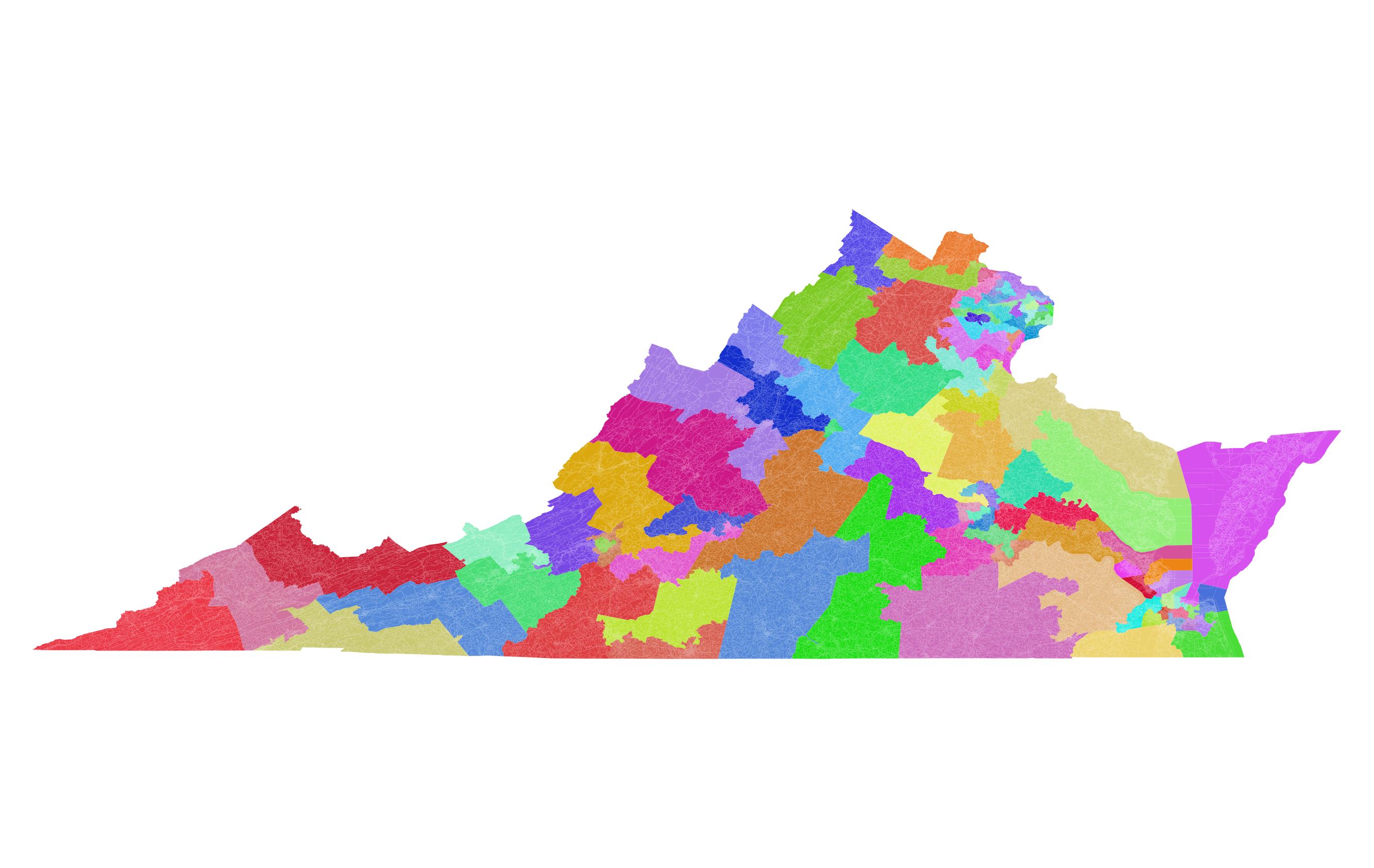}}
        \subfloat[Remedial]{\includegraphics[width=2in]{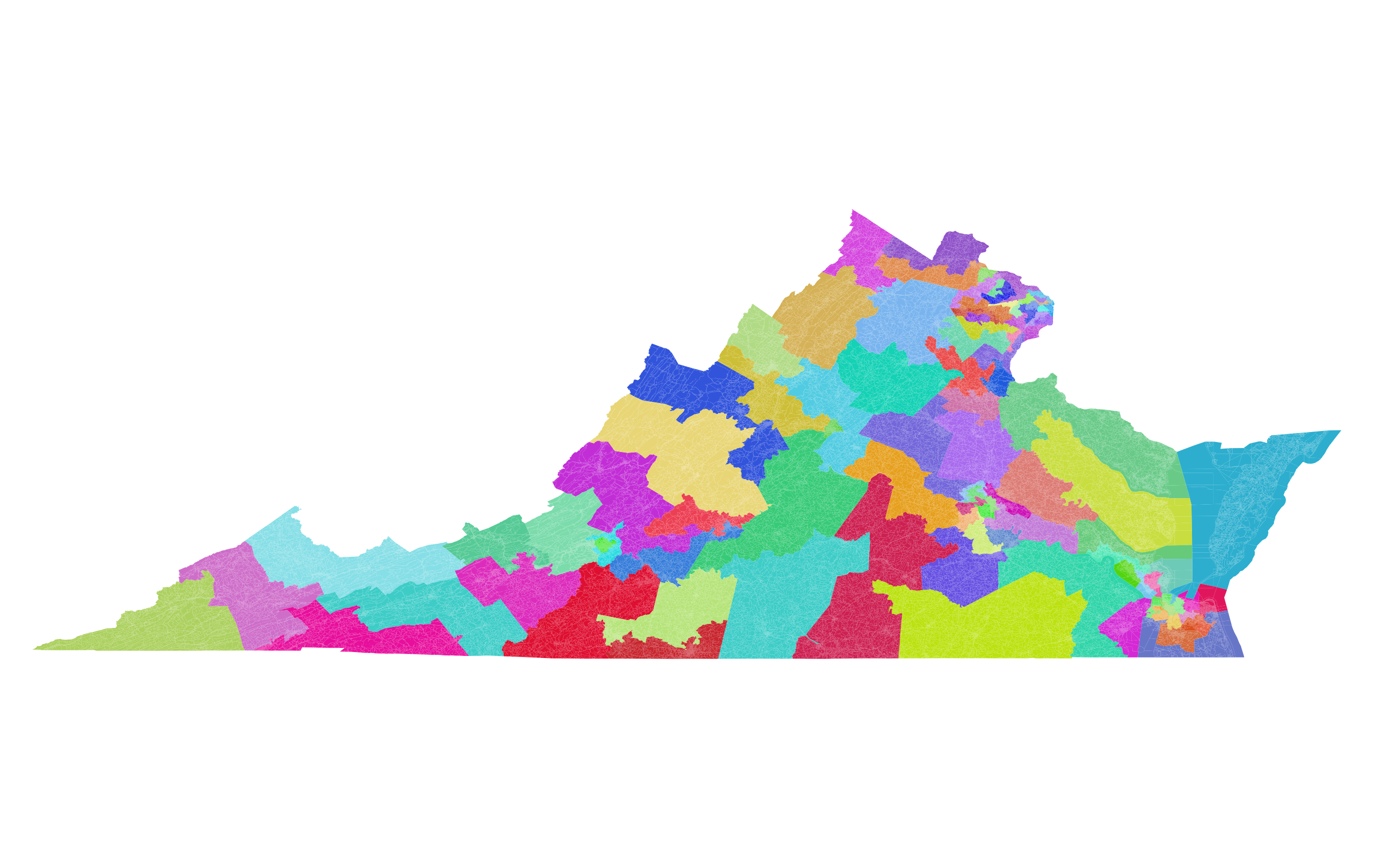}}
        }
        
\begin{tikzpicture}
\node at (0,0) {\includegraphics[width=6in]{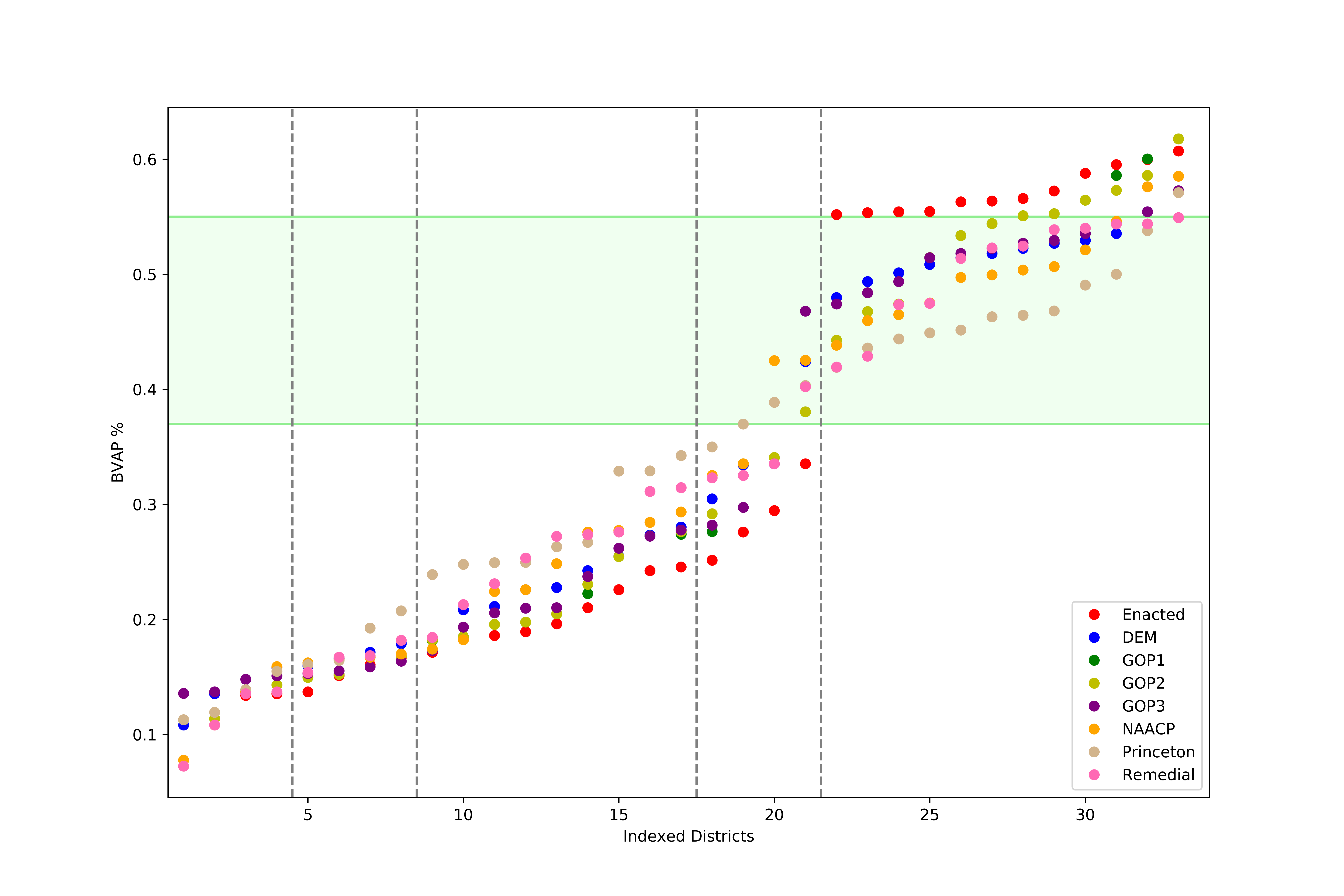}};
\node at (3.9,4.1) {\scriptsize 12 Districts};
\node at (1.02,4.1) {\scriptsize 4 Districts};
\node at (-1.3,4.1) {\scriptsize 9 Districts};
\node at (-3.6,4.1) {\scriptsize 4 Districts};
\node at (-5,4.1) {\scriptsize 4 Districts};
\end{tikzpicture}        
    \caption{Eight proposed House of Delegates plans. 
 The boxplot shows the  Black Voting Age Population in the 33 districts affected by the court ruling, ordered from lowest 
 to highest BVAP in each plan.  
 The 2011 enacted plan jumps the key 37-55\% BVAP range entirely, 
 but the collection of other plans makes it difficult to tell how many more 
 37-55\% BVAP plans  might be expected or possible.}
    \label{fig:proposed_plans}
\end{figure}


\begin{figure}[!h]
    \centering
    \includegraphics[width=4.5in]{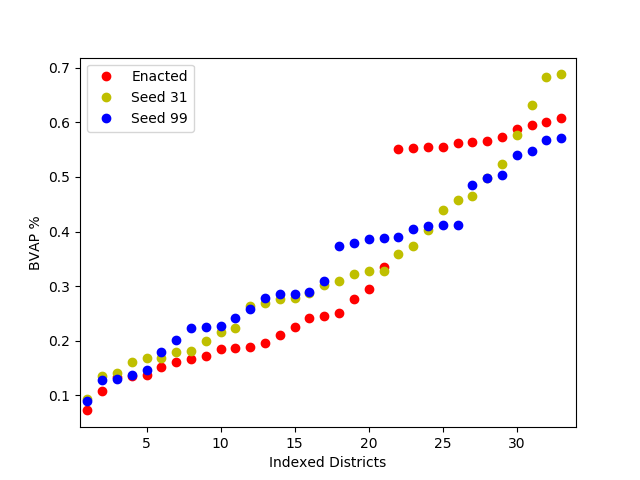}

        \includegraphics[width=3in]{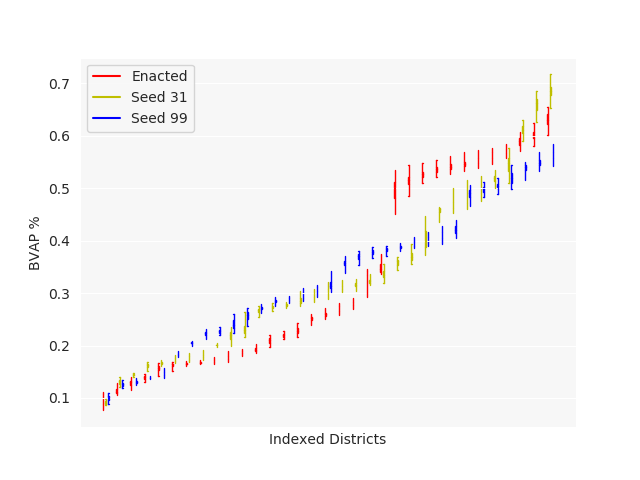}
    \includegraphics[width=3in]{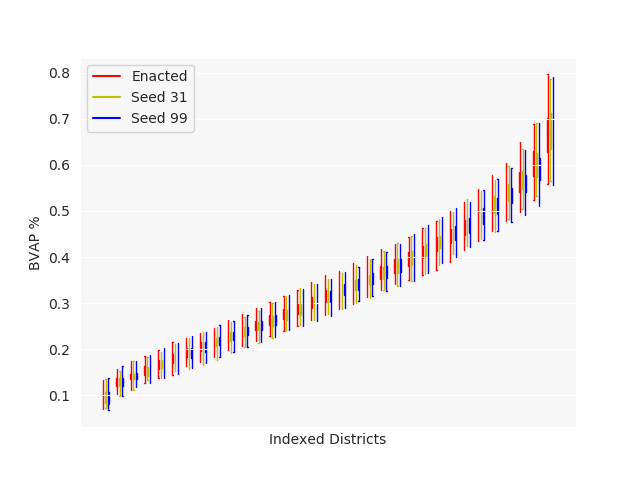}

        \includegraphics[width=3in]{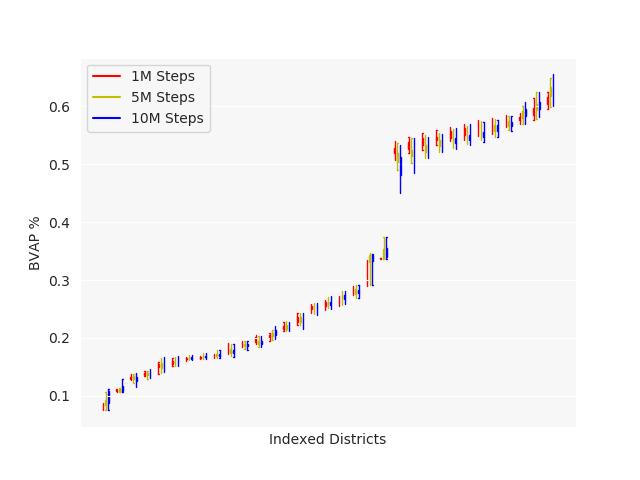}
    \includegraphics[width=3in]{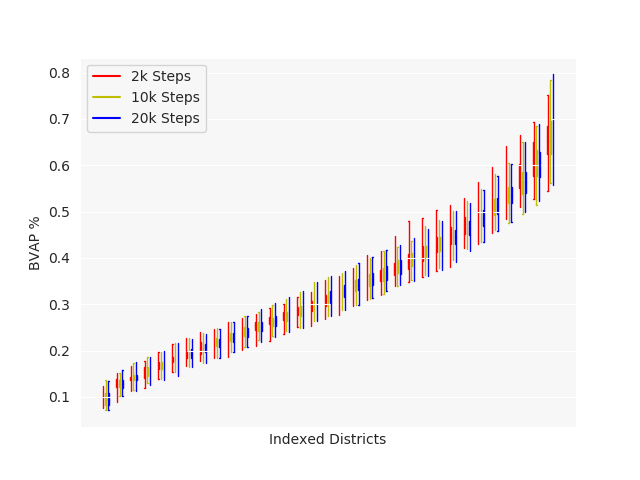}
    \caption{Convergence heuristics.  The BVAP levels in the enacted plan are compared to two synthetically
    generated seed plans.  10 million steps are not enough to mitigate the dependence on the starting point
    in a flip run.  By contrast, 20,000 steps seems to suffice for a recombination run, with most of the progress
    in the first 10,000 steps.  Top row:  levels at starting points.  Middle row: \Flip (left) and \ReCom (right)
    ensembles from three starting points.  Bottom row: runs of varying lengths starting from enacted plan.}
    \label{fig:compare}
\end{figure}

\begin{figure}[!h]
    \centering
\subfloat[\Flip]{\includegraphics[height=1.8in]{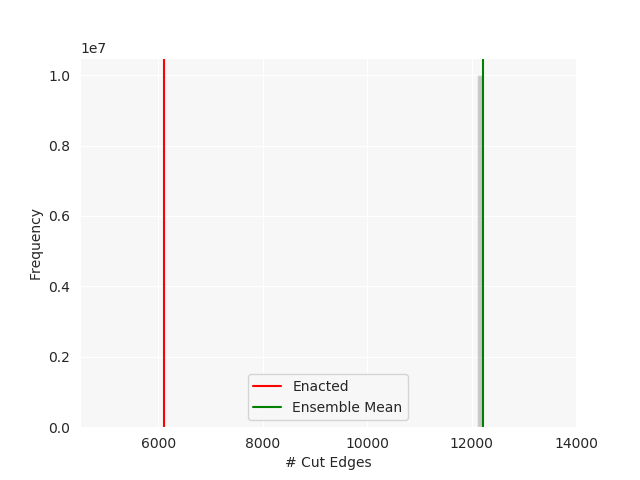}}
\subfloat[\ReCom]{\includegraphics[height=1.8in]{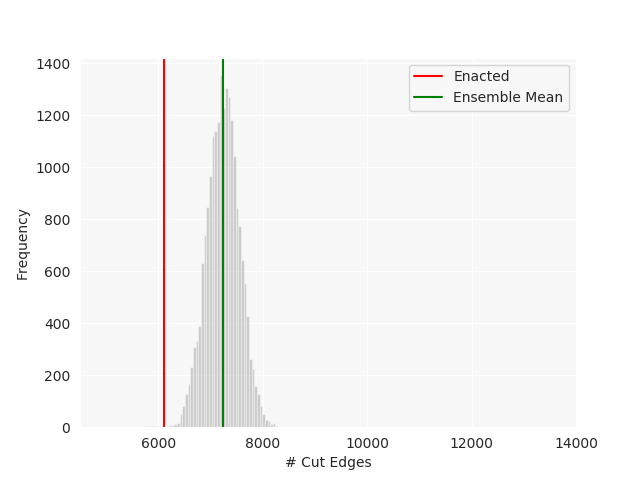}}

\subfloat[\Flip]{\includegraphics[height=1.8in]{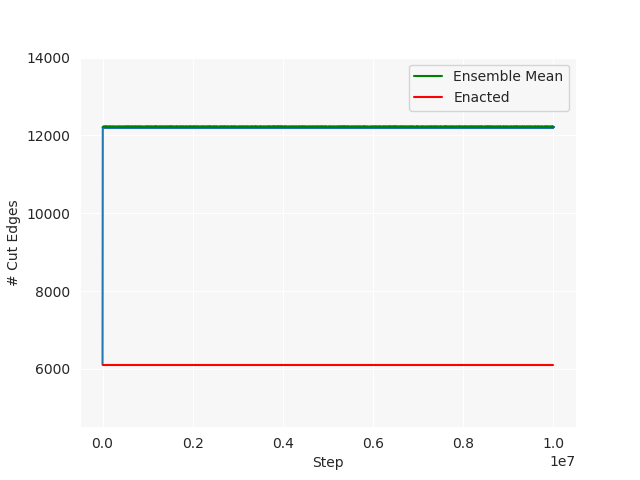}}
\subfloat[\ReCom]{\includegraphics[height=1.8in]{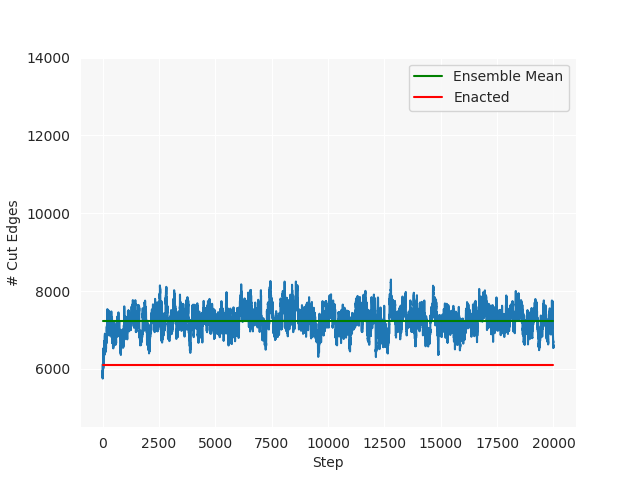}}
    \caption{Compactness comparison. Histograms (a,b) and traces (c,d) of the  boundary length.  Flip ensembles
    saturate the worst allowable compactness score (here, twice the value of the enacted plan).}
    \label{fig:chain_ce}
\end{figure}

\vspace{-.2in}

\begin{figure}[!h]
    \centering
\subfloat[Seed 31]{\includegraphics[height=1.5in]{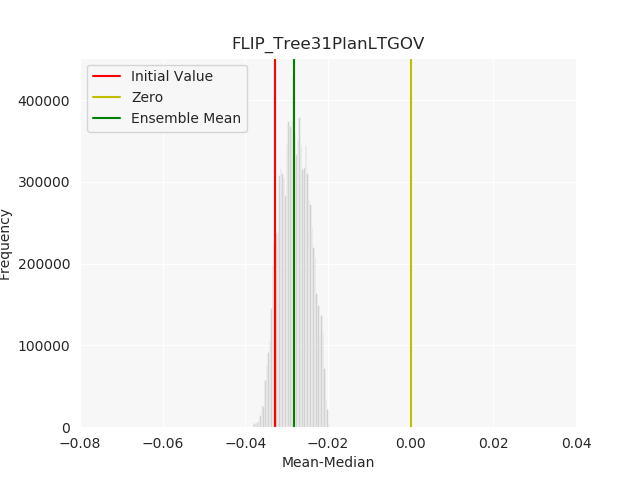}} 
\subfloat[Seed 99]{\includegraphics[height=1.5in]{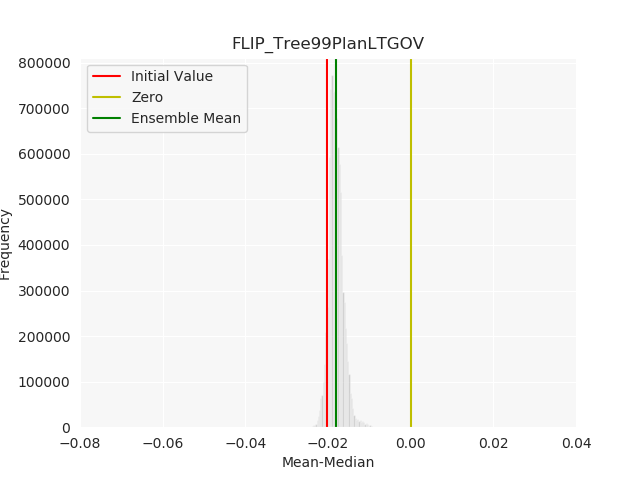}}
\subfloat[Enacted]{\includegraphics[height=1.5in]{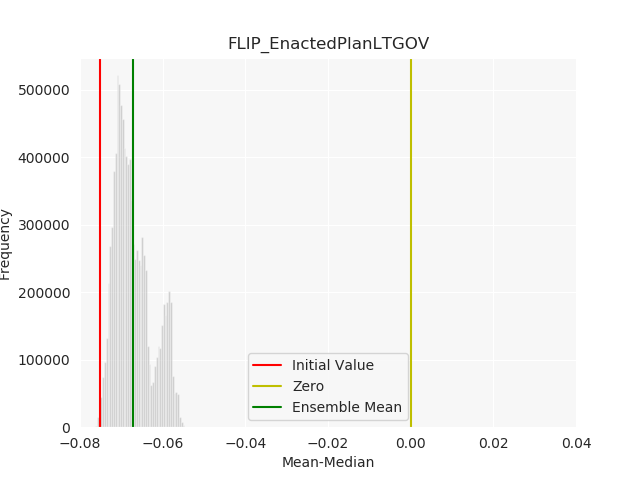}}\\
\subfloat[Seed 31]{\includegraphics[height=1.5in]{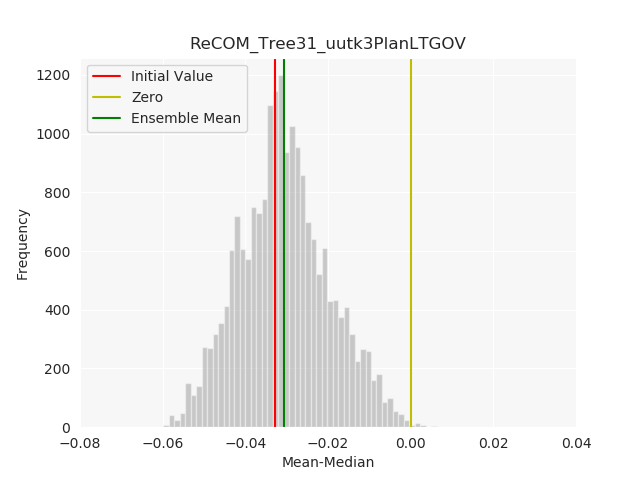}} 
\subfloat[Seed 99]{\includegraphics[height=1.5in]{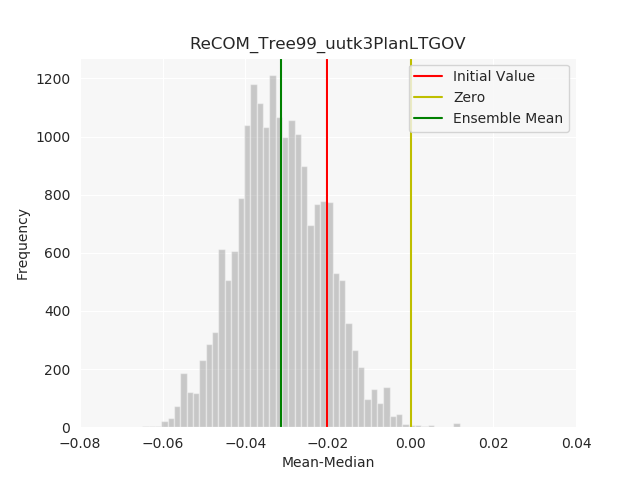}}
\subfloat[Enacted]{\includegraphics[height=1.5in]{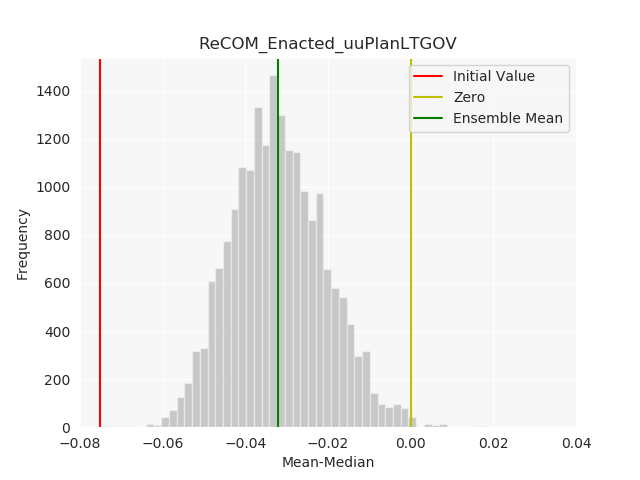}}
\caption{Projection to partisan statistics. Mean-median  (partisan symmetry) scores, illustrating dependence of \Flip ensembles
on starting point after one million steps.}
    \label{fig:mms}
\end{figure}

\begin{figure}[!h]
\centering
\begin{tikzpicture}
\node at (0,0) {\includegraphics[width=6in]{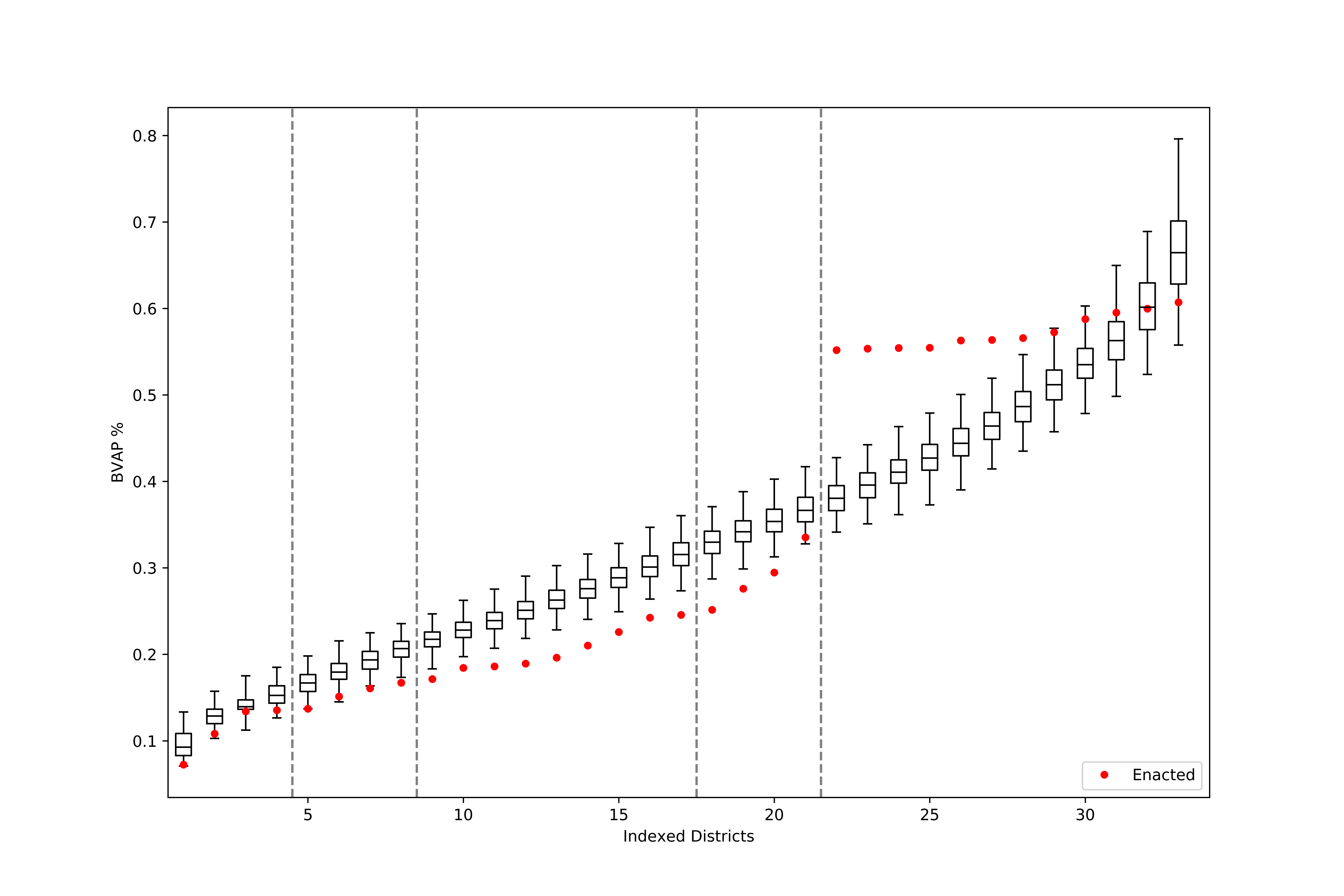}};
\node at (3.9,4.1) {\scriptsize 12 Districts};
\node at (1.02,4.1) {\scriptsize 4 Districts};
\node at (-1.3,4.1) {\scriptsize 9 Districts};
\node at (-3.6,4.1) {\scriptsize 4 Districts};
\node at (-5,4.1) {\scriptsize 4 Districts};
\end{tikzpicture}  
    
\begin{tikzpicture}
\node at (0,0) {\includegraphics[width=6in]{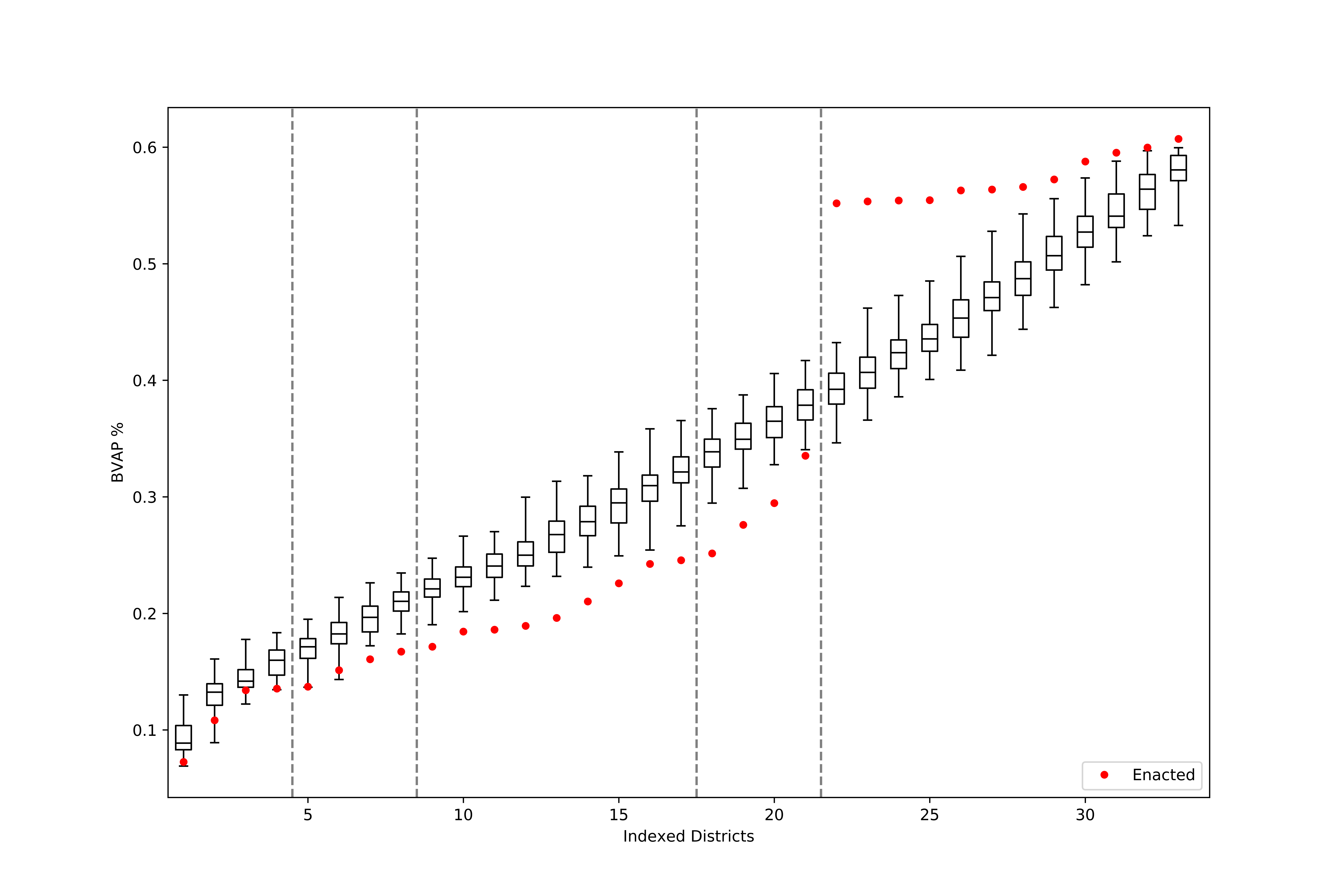}};
\node at (3.9,4.1) {\scriptsize 12 Districts};
\node at (1.02,4.1) {\scriptsize 4 Districts};
\node at (-1.3,4.1) {\scriptsize 9 Districts};
\node at (-3.6,4.1) {\scriptsize 4 Districts};
\node at (-5,4.1) {\scriptsize 4 Districts};
\end{tikzpicture}  
\caption{Ensemble analysis.  The BVAP levels in the Enacted plan can now be compared to an ensemble 
of population-balanced, compact plans that hold the state's demographics and geography constant.
Top: full \ReCom ensemble.  Bottom: same ensemble, winnowed to $\le$60\% BVAP.}
\label{fig:VA-ensembles}
\end{figure}

\end{document}